\documentclass[1p]{elsarticle}

\usepackage{amsfonts}
\usepackage{amsmath}
\usepackage{stmaryrd}
\usepackage{tabularx}
\usepackage{caption}
\usepackage{todonotes}

\usepackage{mymacros}

\journal{Computer Methods in Applied Mechanics and Engineering}

\begin{document}

\begin{frontmatter}

\title{A partition of unity approach to fluid mechanics and fluid-structure interaction }

\author[address1]{Maximilian Balmus~\corref{CA}}
\address[address1]{Department of Biomedical Engineering, School of Imaging Sciences and Biomedical Engineering, King's College London, King's Health Partners, London SE1 7EH, United Kingdom}
\ead{maximilian.balmus@kcl.ac.uk}

%\author{Other Authors}
\author[umu,ntnu]{Andr\'e~Massing}
\ead{andre.massing@ntnu.no}
%\address[address2]{UMIT Reseach Lab, Umea University, Umea, Sweden}
\address[umu]{Department of Mathematics and Mathematical Statistics, Ume{\aa} University, SE-90187 Ume{\aa}, Sweden}
\address[ntnu]{Department of Mathematical Sciences, Norwegian University of Science and Technology, NO-7491 Trondheim, Norway}

\author[kth]{Johan Hoffman}
\ead{jhoffman@kth.se}
\address[kth]{Division of Computational Science and Technology, KTH Royal Institute of Technology, Sweden}

\author[address1]{Reza Razavi}

\author[address1,address3]{David~A.~Nordsletten} 
\cortext[CA]{Corresponding author}
\ead{david.nordsletten@kcl.ac.uk}
\address[address3]{Department of Biomedical Engineering and Cardiac Surgery, University of Michigan, MI, USA}

%\author[mymainaddress,mysecondaryaddress]{Maximilian Balmus}



\begin{abstract}
  For problems involving large deformations of thin structures,
  simulating fluid-structure interaction (FSI) remains challenging
  largely due to the need to balance computational feasibility,
  efficiency, and solution accuracy. Overlapping domain techniques
  have been introduced as a way to combine the fluid-solid mesh
  conformity, seen in moving-mesh methods, without the need for
  mesh smoothing or re-meshing, which is a core characteristic 
  of fixed mesh approaches.
  In this work, we introduce a novel overlapping domain method
  based on a partition of unity approach. Unified function spaces
  are defined as a weighted sum of fields given on two overlapping
  meshes. The method is shown to achieve optimal convergence rates and
  to be stable for steady-state Stokes, Navier-Stokes, and ALE
  Navier-Stokes problems. Finally, we present results for FSI
  in the case of a 2D mock aortic valve simulation. These initial
  results point to the potential applicability of the method to a wide
  range of FSI applications, enabling boundary layer refinement and
  large deformations without the need for re-meshing or user-defined
  stabilization.
\end{abstract}

\begin{keyword}
% \texttt{elsarticle.cls}\sep \LaTeX\sep Elsevier \sep template
%\MSC[2010] 00-01\sep  99-00
Finite element methods \sep Fluid-structure interaction \sep Overlapping domains\sep Partition of unity
\end{keyword}

\end{frontmatter}

%\linenumbers

\newpage

%\amnote{Add missing emails from Max and Reza}
\section{Introduction}
	
Fluid-structure interaction (FSI) problems involving thin solids which undergo large
deformations and translations can be encountered in a significant
number of engineering applications.
In industry, we have examples such as the design of
parachutes~\cite{stein1997parallel,kalro2000parallel,takizawa2011space}
and wind-turbines~\cite{bazilevs20113d,bazilevs20113d2}. In
cardiovascular research, the simulation of
valves~\cite{hsu2015dynamic,gao2017coupled}  and implanted 
devices~\cite{mccormick2014computational}
 holds a great potential
for better understanding and treatment of a number of pathologies such as
valve stenosis, regurgitation, heart failure and outflow obstruction. Building such
models, however, remains challenging due to the need to balance
computational costs and solution accuracy. In the case of cardiac
valves, for example, studies typically require simplifications of the
domain's geometry~\cite{lau2010mitral} or the fluid
models~\cite{baillargeon2015human}.
		
FSI approaches for this class of problems can be
grouped into three main categories: interface-tracking,
interface-capturing and overlapping domains
methods~\cite{bazilevs2013computational,verkaik2014overlapping}. In
the case of interface-tracking, the fluid problem is typically based
on the Arbitrary Lagrangian-Eulerian (ALE)~\cite{hirt1974arbitrary,
donea1982arbitrary, hughes1981lagrangian} formulation. This allows for
the fluid domain to deform with the solid and enables adjusting the
element resolution close to surfaces in order to more accurately
represent boundary layers. In~\cite{van2007comparison},
an ALE based
method is shown to produce superior results when considering moderate
deformations to three variations of fictitious domain (examples of
interface-capturing) for similar mesh resolutions.  However, for large
deformations, it is known that the distortion of the fluid mesh can
diminish the quality of elements and negatively impact the accuracy of
solutions~\cite{bazilevs2013computational}. While multiple re-meshing
techniques~\cite{tezduyar2007modelling} have been proposed, the
process introduces grid interpolation errors and its computational
cost effectiveness is linked to the frequency at which the mesh needs
to be adjusted. As shown in~\cite{bavo2016fluid}, where the particular
case of cardiac valves is discussed, the ALE based simulations take
more time to run than competing interface-capturing methods.

In contrast, interface-capturing methods do not
require boundary fitted meshes for the fluid and solid and thus avoid
the need for re-meshing. Examples include the Fictitious Domain
Method~\cite{glowinski1994fictitious,glowinski1999distributed} (FDM),
where the coupling between the solid and fluid is achieved via
additional Lagrange multiplier terms~\cite{babuvska1973finite}, and
the Immersed Finite Element method (IFEM)~\cite{zhang2004immersed,ZhangGay2007,BoffiCavalliniGastaldi2015,BoffiGastaldi2017},
where the kinematic constraints between the fluid and solid is imposed
through interpolation and distribution of local body forces. While
both FDM and IFEM require the construction of a solid mesh, in the
Immersed Structural Potential Method (ISPM)~\cite{gil2010immersed}
both problems are solved on the same mesh, with the solid being
represented as a moving collection of quadrature points. However,
these methods lack a conforming interface between the fluid and solid
and can result in poor approximations of pressure jumps and surface
stresses~\cite{dos2008partitioned,kamensky2015immersogeometric}. For
this reason, mesh-adaptation~\cite{van2004combined} and XFEM
enrichment~\cite{gerstenberger2008extended,alauzet2016nitsche,Landajuela2016,Schott2017,MassingSchottWall2017,WinterSchottMassingEtAl2017,ZoncaFormaggiaVergara2016,FormaggiaVergaraZonca2018} techniques have
been proposed in order to overcome these issues.
Also a significant
challenge for interface-capturing, as noted
in~\cite{tezduyar2003computation}, remains the fact the resolution of
the fluid flow around the structural surface is limited by the local
element size of the fluid mesh. In practice, this leads to
over-refinement of the fluid mesh along the moving trajectory of the
solid and to significant increases of the computational costs.
		
Overlapping domain
techniques~\cite{StegerDoughertyBenek1983, StegerBenek1987,ChesshireHenshaw1990,HouzeauxCodina2003,WallGamnitzerGerstenberger2008,hansbo2003finite,MassingLarsonLoggEtAl2015,SchottShahmiriKruseEtAl2016,verkaik2014overlapping,KoblitzLovettNikiforakisEtAl2017}
% techniques~\cite{StegerDoughertyBenek1983,ChesshireHenshaw1990,HouzeauxCodina2003,WallGamnitzerGerstenberger2008, hansbo2003finite,MassingLarsonLoggEtAl2015,SchottShahmiriKruseEtAl2016,verkaik2014overlapping}
have been proposed with the aim of combining the advantages of
interface-tracking and interface capturing techniques: fluid-solid
mesh conformity, boundary layer tracking and eliminating re-meshing.
The crux of these methods is the decomposition of the fluid problem
into a background coarse component and solution-enriching embedded
component that envelops the structures. A challenging aspect, however,
is the coupling of the two fluid domains which has to be done weakly
due the non-matching fluid-fluid interface. The use of Lagrange
multipliers for example, see~\cite{gerstenberger2008extended}, is
impeded by the need to properly choose function spaces in order to
guarantee that the inf-sup condition holds for arbitrary moving
interfaces. Alternatively, stabilization techniques can be employed
to circumvent the inf-sup
condition~\cite{Burman2013,BurmanHansbo2010,PusoKokkoSettgastEtAl2014}.  Overlapping mesh
methods for the Stokes
problem~\cite{massing2014stabilized,JohanssonLarsonLogg2015,SchottShahmiriKruseEtAl2016} which use
Nitsche's method~\cite{nitschevariationsprinzip} avoid introducing
an additional Lagrange multiplier field, but
nevertheless they require additional, parameter-dependent stabilization
terms for the velocity and pressure jump in the vicinity of the
fluid-fluid interface to guarantee optimal convergence rates and good
system conditioning irrespective of the particular overlap
configuration.

In this paper, we propose a new flexible and robust overlapping domain
method which uses the partition of unity (PUFEM)
approach~\cite{melenk1996partition,HuangXu2003} to decompose the fluid domain into
a background mesh and an embedded mesh which can overlap in an arbitrary manner.
On each mesh,
standard mixed and inf-sup stable velocity-pressure function spaces using
Taylor-Hood elements are defined.
In the final finite element formulation of the fluid problem, 
a unified global function space is then used which is defined by
taking a properly weighted sum of the function spaces associated with each mesh.
To avoid ill-conditioning of the resulting system, additional constraints are
introduced in (parts of) the overlap region.

The rest of this paper is structured as follows.  After briefly
recalling the classical mixed finite element approach for the Stokes
problem in Section~\ref{sec:classic_Stokes}, we introduce its PUFEM
based overlapping domain formulation including a detailed description
of the domain set-up, weighted function spaces, and imposed
constraints, see Section~\ref{sec:PUFEM_Stokes}, followed by a short
discussion of some computational aspects in
Section~\ref{sec:alg_intersection}.  Then in
Section~\ref{sec:PUFEM_4_ALE_and_FSI}, we explain how to combine the PUFEM
approach with an ALE formulation of the Navier-Stokes problem to treat
moving fluid domain and fluid--structure interaction problems.  In
Section~\ref{sec:results}, a number of numerical experiments are
conducted.  First, we investigate the stability and accuracy of our PU
approach for a number of fluid flow problems posed on fixed static
domains, see
Section~\ref{sec:res_stokes}--\ref{ssec:schaf-turek-benchm}.  To
demonstrate the capability to handle large changes in the fluid domain
geometry, we then consider a fluid flow driven by a oscillating
cylinder in Section~\ref{ssec:ale-oscill-cyl} before we turn to a full
FSI problem in Section~\ref{sec:valve-simulation}, where we compare a
classical ALE-based approach with our novel combined ALE-PUFEM
discretization for a two-dimensional mock aortic valve simulation.
Finally in Section~\ref{sec:conclusion}, we summarize our results and
discuss potential future developments.
%\amnote{Remember to adapt this outline if we decide to split Section 2 into two sections.}
%\mbnote{Section has been split as it was discussed.}

% Here we outline the
% key aspects of the technique and show the approach produces optimal
% convergence over a range of fluid test problems without the need for
% stabilization terms and user defined parameters. Our main goal is to
% show that PUFEM can be used to achieve comparable levels of solution
% accuracy with the classical ALE approach, which serves as our gold
% standard.
		
% The rest of this paper is structured as follows. In Section 2, we
% present the PUFEM technique in the context of the Stokes problem,
% incompressible Navier-Stokes in ALE form and FSI. In Section 3 we show
% that the method is accurate and stable for a range quasi-static,
% transient and ALE flow problems as well as for a demonstrative FSI
% simulation. In the last section, we will discuss the main conclusions
% as well as potential future developments.

%%% Local Variables:
%%% mode: latex
%%% TeX-master: "Max_Article"
%%% End:

\section{A Partition of unity finite element method for the Stokes problem}
\label{sec:method}
 
In this section, we introduce the main concepts 
and the basic setup for PUFEM.  
We begin by reviewing the classic mixed FEM
approach to Stokes flow (Section~\ref{sec:classic_Stokes}) and present
the key differences introduced in the PUFEM approach (Section
~\ref{sec:PUFEM_Stokes}). Finally, in Section~\ref{sec:alg_intersection}, we
describe the process by which we identify the polygonal intersections
of overlapping elements and perform the necessary
integrations. 

\subsection{Classical mixed FEM approach to  Stokes problems}
\label{sec:classic_Stokes}

Let $\Omega_f\subset \realset^d$ be an arbitrary fluid domain
on which we solve our problem. Since our focus is oriented towards FSI
simulations, we also introduce a solid domain $\Omega_s$ which for
now, we consider to be fixed and rigid. The boundary of the fluid
problem, $\Gamma$, is composed of three, non-overlapping regions: the
portions where Dirichlet and Neumann boundary conditions are applied,
($\Gamma_{D}$ and $\Gamma_{N}$, respectively), and the fluid-solid
interface, $\Gamma_{fs}$.
We then can write Stokes problem as: find the $(\bv,p)$ such that,
\begin{subequations}
  \begin{align}
    \mu \nabla^2 \bv - \nabla p &= \mathbf{0} && \text{ in } \Omega_f, \\
    \divergence \bv & = 0 && \text{ in } \Omega_f, \\
    \left(\mu\nabla\bv - p\bI  \right) \cdot \bn & =\bt_N && \text{ on }  \Gamma_N, \\
    \bv & = \bv_D && \text{ on } \Gamma_D , \\
    \bv & = \mathbf{0} && \text{ on } \Gamma_\mathit{fs},
  \end{align}
  \label{eq:Stokes_strong}
\end{subequations}
where $\mu$ is the fluid dynamic viscosity constant. For simplicity, we
consider that the Neumann boundary condition tractions are null, 
i.e. $\bt_N = \mathbf{0}$.
	
In the case of the Stokes problem in (\ref{eq:Stokes_strong}),
the classic continuous weak form can be written as: find
$(\bv,p)\in\boldsymbol{\mathV}_D\times\mathW$ such that,
\begin{equation}
  \int_{\Omega} \mu\nabla\bv : \nabla \bw - p\divergence\bw + q\divergence \bv\hspace{1mm} d\Omega = 0, 
  \hspace{5mm} \forall\hspace{1mm} (\bw,q)\in\boldsymbol{\mathV}_0\times\mathW,
\end{equation}
where $\boldsymbol{\mathV}_D$,
$\boldsymbol{\mathV}_0\subset\bH^1(\Omega_f)$ and
$\mathW=L^2(\Omega_f)$.
Here the subscripts $D$ and $0$ indicate that
the $\boldsymbol{\mathV}_D$,
$\boldsymbol{\mathV}_0 \subset \boldsymbol{\mathV}$ subspaces are
built such that they incorporate the Dirichlet and zero boundary value
conditions on $\Gamma_{D}$.  Proofs of the well-posedness of this
problem can be found in~\cite{quarteroni1994introduction}
and~\cite{girault2012finite}.  In the discrete setting, the resulting
weak form is: find
$(\bv^h,p^h)\in\boldsymbol{\mathV}^h_D\times\mathW^h$ such that,
\begin{equation}
  \label{eq:Stokes_weak}
  \int_{\Omega^h_f} \mu\nabla\bv^h : \nabla \bw^h - p^h\divergence\bw^h + q^h\divergence \bv^h\hspace{1mm} d\Omega = 0,
  \hspace{5mm} \forall \hspace{1mm} (\bw^h,q^h)\in \boldsymbol{\mathV}^h_0\times\mathW^h.
\end{equation}

In this study, we will use the classic approach
in~(\ref{eq:Stokes_weak}) to compare with the PUFEM approach. In this
case, we use the LBB stable $\polyset^2-\polyset^1$ Taylor-Hood
elements~\cite{taylor1973numerical}. Thus, the discrete test and trial
function spaces can be defined as:
\begin{subequations}
  \begin{align}
    \boldsymbol{\mathV}^h & = \{\bv^h \in \bC^0 (\Omega^h_f) \defas \bv|_\tau \in\left[\polyset^2(\tau) \right]^d \text{ for } \tau\in\mathcal{T}_f  \}, \\
    \mathW^h & = \{ p^h \in C^0(\Omega^h_f) \defas p|_\tau \in \polyset^1(\tau) \text{ for } \tau\in\mathcal{T}_f  \} .
  \end{align}
\end{subequations}
Here $\Omega^h_f$ is a discrete equivalent of $\Omega_f$, defined 
by a tessellation $\mathcal{T}_f$. $h$ denotes the element
size defined as the diameter of the circumcircle. In preparation for the PUFEM
discussion, we can also define  $\Omega^h_s$ as the discrete solid
domain. $\polyset^k$
designates the set of polynomial function of order $k$.

Previous works~\cite{quarteroni1994introduction,girault2012finite}
derive \textit{a priori} error estimates where:
\begin{align}
  ||\bv - \bv^h||_1 + ||p - p^h||_0  & \leq C_1 \inf_{\bw^h\in\boldsymbol{\mathV}^h}  ||\bv - \bw^h||_1 +  C_2  \inf_{q^h\in\mathW^h} ||p - q^h||_0 \label{eq:ineq1},
\end{align}
where $C_1$ and $C_2$ are positive constants independent of $h$, and
$||\cdot||_0$ and $||\cdot||_1$ denote the $L^2$ and $\bH^1$ norms,
respectively. Using $\polyset^2-\polyset^1$ elements, if
$(\bv, p)\in \bH^3(\Omega) \times H^2(\Omega)$, then from
interpolation theory~\cite{quarteroni1994introduction}:
\begin{align}
  ||\bv - \bv^h||_1 + ||p - p^h||_0  \leq Ch^2\left(|\bv|_3  +|p|_2 \right). \label{eq:ineq5}
\end{align}	 

\subsection{PUFEM approach to Stokes problems}
\label{sec:PUFEM_Stokes}
% \amnote{Remember to look into/refer to~\cite{HuangXu2003}
%   where they considered a very similar a PUM for elliptic PDEs
%   on overlapping meshes, including some numerical analysis.
%   For conditioning, \cite{Li2012} might be relevant.
% }

\begin{figure}[!]
	\centering
	\includegraphics[trim={0cm 1cm 0 2cm},clip,page=2,width=\textwidth]{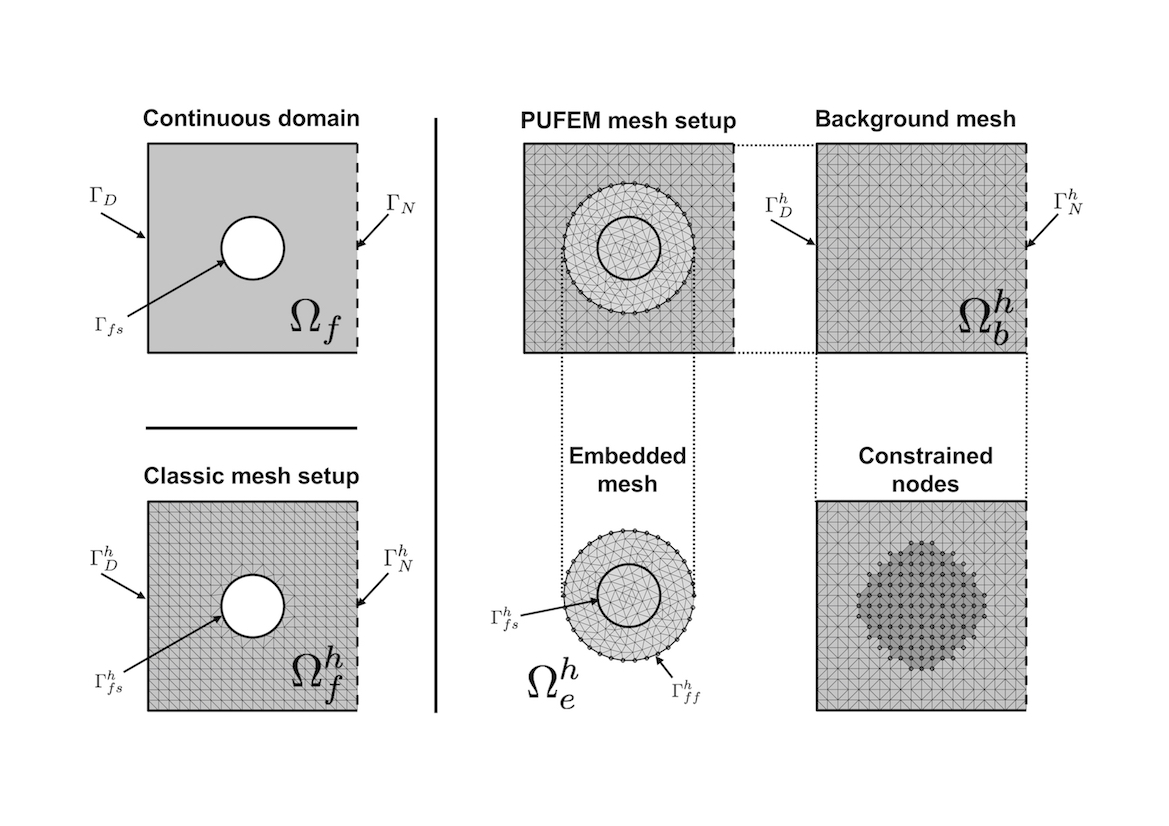}
	\caption{The problem domain in three instances: continuous, classical, single mesh setup and PUFEM setup. In the later, the discrete domain is subdivided into the embedded and background overlapping meshes. Both meshes contain a set of constrained nodes used to avoid ill-conditioning. Here we mark them with circles.  }
	\label{fig:PUFEM_domain}
\end{figure}

The core difference between the classic approach and the PUFEM setup
is that the latter is composed of two overlapping meshes: background
and embedded (see Fig.~\ref{fig:PUFEM_domain}), and both have a 
corresponding discrete domain over which they are defined.
Thus, we assume the
background domain, $\Omega^h_b = \Omega^h_f \cup \Omega^h_s$,
encompasses the entirety of the discrete fluid and solid domains. The
embedded domain, $\Omega^h_e$, which satisfies
$\Omega^h_s \subset \Omega^h_e \subseteq \Omega^h_b$, is designed to
incorporate the solid and extends into fluid domains providing a
boundary layer. The fluid and solid regions of $\Omega^h_e$ are
separated by the $\Gamma^h_\mathit{fs}$ interface. Additionally,
$\Gamma_{\mathit{ff}}^h$ is the outer boundary of $\Omega^h_{e}$ and
serves as its interface with $\Omega^h_{b}$. The two meshes each have a
corresponding tessellation, $\mathcal{T}_{b}$ and $\mathcal{T}_{e}$,
and element sizes denoted by $h_{b}$ and $h_{e}$, respectively. Let
$N^v_b$ and $N^p_b$ be the set of nodes defining the background mesh
in the case of quadratic and linear interpolations. $N^v_e$ and
$N^p_e$ are their embedded analogues. For the remainder of this paper,
we consider the case where the meshes are 2D and triangular.

The discrete weak form for the PUFEM approach can be written as: find
$(\bv^h,p^h)\in\bV^h_D\times W^h$ such that,
\begin{equation} \int_{\Omega^h_b} \mu\nabla\bv^h : \nabla \bw^h -
p^h\divergence\bw^h + q^h\divergence \bv^h\hspace{1mm} d\Omega = 0,
\hspace{5mm} \forall \hspace{1mm} (\bw^h, q^h)\in\bV^h_0\times W^h
  \label{eq:PUFEM_Stokes_weak_form}
\end{equation}
where $\bV^h_D$, $\bV^h_0$ and $W^h$ are PUFEM function
spaces that represent the counterparts of $\boldsymbol{\mathV}_D^h$,
$\boldsymbol{\mathV}_0^h$ and $\mathW^h$ introduced in
Section~\ref{sec:classic_Stokes}.  The PUFEM function spaces are
defined as the weighted sums of function spaces with support on the
background and embedded meshes. Thus, based on the same notation of
Melenk and Babuska~\cite{melenk1996partition}, the spaces can be
written as:
\begin{subequations}
  \begin{align}
    \bV^h & = \left(1- \psi^h\right)\bV^h_{{b},*} + \psi^h\bV^h_{{e},*} \\
    W^h  & = \left(1 - \psi^h\right) W^h_{{b},*} + \psi^h W^h_{{e},*}
  \end{align}
  \label{eq:PUFEM_spaces}
\end{subequations}
where the local spaces are defined as:
\begin{subequations}
  \begin{align}
    \bV^h_{{k},*}  & = \{ \bv_k^h \in \bV^h_{k} \defas  \bv_k|_{\bx} = \bv_m|_{\bx}	 \text{ for } \bx\in\bX^v_k \} \\
    W^h_{k,*} & = \{ p_k^h \in W^h_k \defas p_k|_{\bx} = p_m|_{\bx} \text{ for } \bx \in \bX^p_k \}
  \end{align}
  \label{eq:local_spaces_w_bc}
\end{subequations}
for $k,m\in\{b,e\}$, $k\neq m$, and
\begin{subequations}
  \begin{align}
    \bV^h_{{k}} & = \{  \bv^h \in \bC^0(\Omega^h_k) \defas \bv^h|_\tau \in \left[\polyset^2(\tau)\right]^d \hspace{1mm} \forall \tau\in\mathcal{T}_k \} \text{ for } k\in \{b,e\} \\
    W^h_b & = \{ p^h\in C^0(\Omega^h_g) \defas p^h|_\tau \in \polyset^1(\tau) \hspace{1mm} \forall \tau \in \mathcal{T}_b \} \\
    W^h_e & = \{ p^h\in L^2(\Omega^h_e) \cap \left[C^0(\Omega^h_s) \oplus C^0(\Omega^h_e\cap\Omega^h_f)\right] \defas p^h|_\tau \in \polyset^1(\tau) \hspace{1mm} \forall \tau \in \mathcal{T}_e  \}
  \end{align}	
  \label{eq:local_spaces}
\end{subequations}
Note that with the exception of $W^h_e$, all the discrete local spaces
are continuous.

We allow the functions in $W^h_e$ to be discontinuous
across $\Gamma^h_\mathit{fs}$ in order to be able to represent the
pressure jump between the fluid and embedded solid. The $\bX^v_k$ and
$\bX^p_k$ sets contain nodes that are constrained in order to avoid
ill-conditioning. More details on how these sets are constructed can
be found in Section~\ref{sec:Redundant_DOF}. 

\begin{figure}[!]
	\centering
	\includegraphics[trim={0cm 0cm 0 3cm},clip,width=\textwidth]{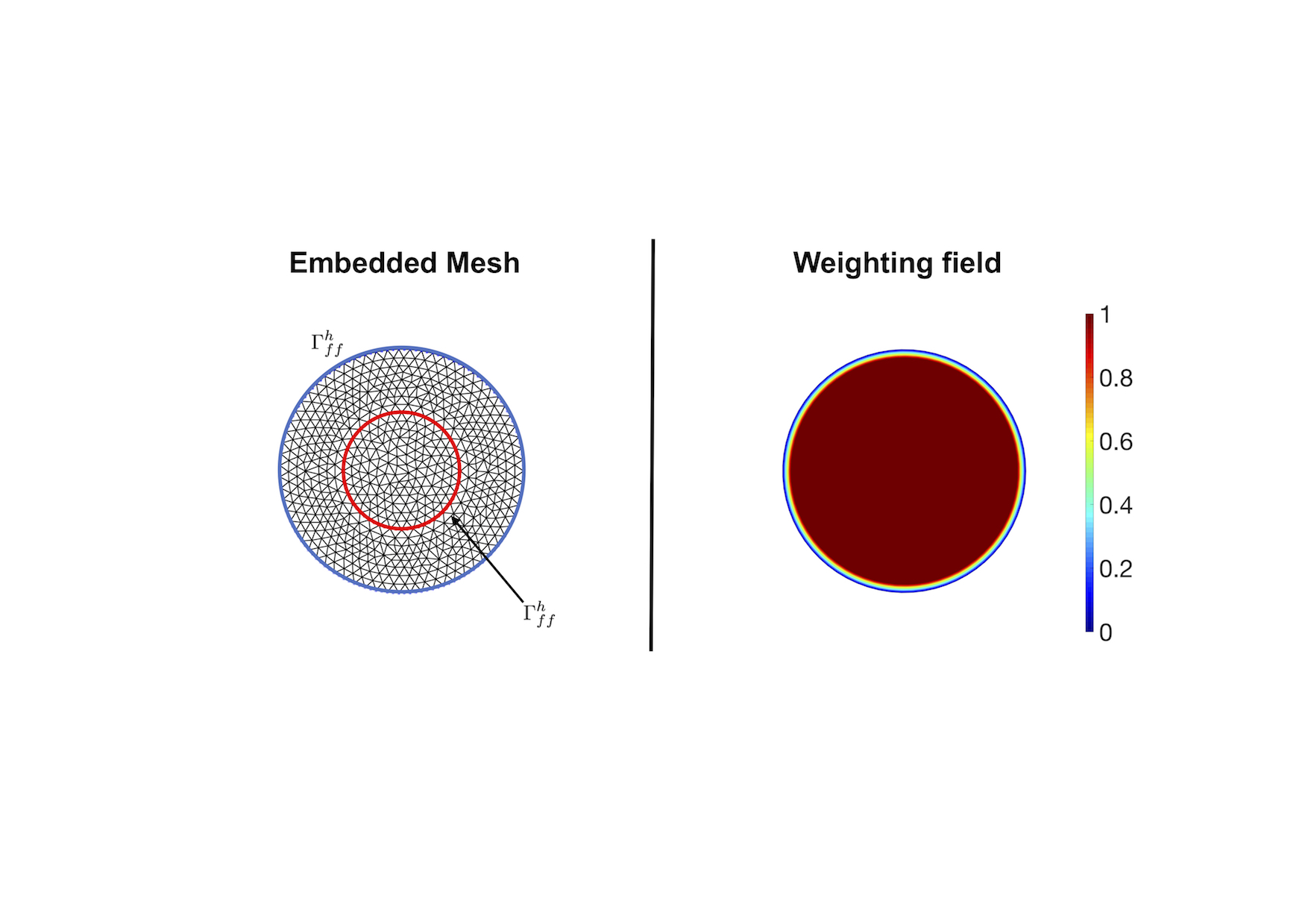}
	\caption{ (Left) An example of the embedded mesh with marked fluid-fluid and fluid-solid interfaces. (Right) 
          The resulting weighting field ($\psi^h$) }
	\label{fig:Stokes_psi}
      \end{figure}

We define the field $\psi^h\in V^h_e$ ($V^h_e$ is the scalar equivalent of $\bV^h_e$) such
that $\supp(\psi^h)\subset\Omega^h_e$, its codomain is $[0,1]$ and
$\psi^h = 0$ on $\Gamma_\mathit{ff}^h$. For notational simplicity, we
consider that all fields defined on the embedded mesh (i.e. $\psi^h$,
$\bv^h_e$ and $p^h_e$) are zero outside the region of overlap. 
In practice, we built $\psi^h$ using the following series
of steps. First, we solved a diffusion problem on the embedded mesh, where we constrain the 
field to be zero on $\Gamma_\mathit{ff}^h$ and ten on $\Gamma_\mathit{fs}^h$. 
Subsequently, we apply a upper boundary threshold such that the maximum node value is one.
Finally, we smoothen the field using the hermitian polynomial $f(u)=-2u^3 + 3u^2$.
An example of the resulting field in the case of circular mesh can be seen Fig.~\ref{fig:Stokes_psi}.
This approach was used irrespective of element size for consistency.
Note however, that this process was chosen on an ad hoc basis. A full 
investigation on the effects of the support area and shape of $\psi$ remain to 
be investigated.

One of the benefits that we derive from the weighted form of the PUFEM is that
it guarantees a smooth transition across $\Gamma_\mathit{ff}^h$, irrespective 
of the mesh size used in either background or embedded mesh. 
As a result, this should always prevent
having mismatched fluxes (i.e. when computed on both sides of the boundary) 
and the loss of mass across the interface.

\subsubsection{Potential sources of ill-conditioning and constraints}
\label{sec:Redundant_DOF}

Previously we introduced four sets of nodes that we want to constrain
$\bX^v_b$, $\bX^v_e$, $\bX^p_b$ and $\bX^p_e$ and in this subsection
we will discuss both their necessity and how we decide which nodes
belong to them. With the introduction of the weighting field, the
standard basis function support concept becomes insufficient to
understand the contribution of each DOF to the total solution. Thus,
we introduce a new metric of effective support fraction which for an
embedded DOF we define as:
\begin{equation}
  E(\bx_i) = \left[\int_{\Omega_f^h} (\psi^h \hat{\phi}^i)^2d\Omega\right] \left[\int_{\Omega_f^h} (\hat{\phi}^i)^2d\Omega\right]^{-1}
\end{equation}
where $\hat{\phi}^i$ is the basis function of a degree of freedom
corresponding to the $\bx_i$ node coordinate. Note that in the case of
a background DOF we replace $\psi^h$ with $1-\psi^h$.
As
$E(\bx_i)\rightarrow 0$ due to a particular overlap configuration, the
contribution of DOFs corresponding to $\bx_i$ become very small. If
$E = 0$ then the contribution to the PUFEM solution is null and the
system matrix becomes singular. Thus, we define the four sets of
constrained nodes as follows:
\begin{subequations}
  \begin{align}
    \bX^v_e & = \{\bx_i\in N^v_e\defas \bx_i\in \Gamma^h_{\mathit{ff}} \} \label{eq:constrianed1}, \\
    \bX^p_e & = \{\bx_i\in N^p_e\defas\bx_i\in \Gamma^h_{\mathit{ff}} \label{eq:constrianed2}  \} ,\\
    \bX^v_b & = \{\bx_i\in N^v_b\defas\bx_i\in \bar{\Omega}^h_{cut}\text{, } \bx_i\notin \Gamma^h_{\mathit{ff}} \label{eq:constrianed3}  \}, \\
    \bX^p_b & = \{\bx_i\in N^p_b\defas\ E(\bx_i) < \epsilon \} \label{eq:constrianed4} .
  \end{align}
\end{subequations}

In the case of the constraints on $\Gamma_{\mathit{ff}}$,
i.e.~(\ref{eq:constrianed1}) and~(\ref{eq:constrianed2}), we have not
observed particularly small values of the $E$ metric. However, we
decided to include these nodes in order to obtain a smoother solution
for the embedded mesh. $\bar{\Omega}^h_{cut}$ refers to the area of
the background domain comprised of elements completely covered by the
embedded mesh. The definition in~(\ref{eq:constrianed3}) is meant to
prevent any ill-conditioning, due to having a non-unique solution. The
second condition, that the background nodes do not lie on fluid-fluid interface,
 is meant to prevent a circular definition (e.g. $\bv_b(\bx_i) = \bv_e(\bx_i)$, 
 therefore $\bv_e(\bx_i) = \bv_b(\bx_i)$). In the case of~(\ref{eq:constrianed4}), we chose to reduce
the number of constrained DOF compared to $\bX^v_b$ and generally assigned
$\epsilon$ to be 0.1. In tests which are not shown here, we saw that increasing
the number of constrained pressure nodes can lead to a deterioration of
the incompressibility condition at the domain level. While this effect can 
be lowered by decreasing $\epsilon$, the trade off is an increase in
the system's condition number. 

%%% Local Variables:
%%% mode: latex
%%% TeX-master: "Max_Article"
%%% End:

\subsubsection{Fully discrete form of PUFEM for Stokes problems}
Following the definition in
Eq.~(\ref{eq:PUFEM_spaces}),~(\ref{eq:local_spaces})
and~(\ref{eq:local_spaces_w_bc}), the discrete fields $\bv^h$ and
$p^h$ can be expanded into the sum of weighted basis functions:
\begin{subequations}
  \label{eq:discrete-function-as-weighted-sum}
  \begin{align}
    \bv^h(\bx) &= \sum_I \left[1-\psi^h(\bx)\right] \hat{\phi}_v^I(\bx) \tilde{\bv}^I_b  +\sum_J  \psi^h(\bx) \hat{\zeta}^J_v(\bx) \tilde{\bv}^J_e \label{eq:PUFEM_weighted_sum}, \\
    p^h(\bx) & = \sum_K \left[1-\psi^h(\bx)\right]  \hat{\phi}_p^K(\bx)\tilde{p}^K_b  + \sum_L \psi^h(\bx)\hat{\zeta}^L_p(\bx) \tilde{p}^L_e.
  \end{align}
\end{subequations}
$\hat{\phi}$ and $\hat{\zeta}$ denote the piecewise polynomial
functions defined on global and embedded meshes, respectively. Their
subscripts are used in order to differentiate between first order ($p$)
and second order ($v$) functions, while the superscript marks the
degree of freedom (DOF) index. Thus, the four sets of basis functions
can be written as: $\{\hat{\phi^I_v}\}_{1\leq I \leq N_{v,b}}$,
$\{\hat{\phi}^K_p\}_{1\leq K\leq N_{p,b}}$,
$\{ \hat{\zeta}^J_v \}_{1\leq J \leq N_{v,e}}$ and
$\{ \hat{\zeta}^L_p \}_{1 \leq L \leq N_{p,e}}$. $N_{v,b}$, $N_{v,e}$,
$N_{p,b}$ and $N_{p,e}$ are the total numbers of DOFs corresponding to
the $\bV^h_b$, $\bV^h_e$, $W^h_b$ and $W^h_e$ spaces. The nodal DOF
vectors $\tilde{\bv}^I_b$ and $\tilde{\bv}^J_e$ contain $d$ entries
each (e.g.
$\tilde{\bv}^I_b = \left[ v^I_{b,1},\cdots, v^I_{b,d} \right]^T$).
Note that, by design, this definition varies according to
where we evaluate the field. Thus, for $\bx \in \Omega^h_e$ we have
the definition in Equation~(\ref{eq:PUFEM_weighted_sum}), while for
$\bx\in\Omega^h_b\backslash\Omega^h_e$ the expansion simplifies to:
\begin{equation*}
  \bv^h(\bx) = \sum_I\hat{\phi}_v^I(\bx) \tilde{\bv}^I_b \hspace{.5cm} \text{ and } \hspace{.5cm}	p^h(\bx)  = \sum_K \hat{\phi}_p^K(\bx)\tilde{p}^K_b . 
\end{equation*}

We can re-arrange the
DOFs into four distinct vectors of unknown. For the background mesh, for
example, we have:
\begin{align*}
  \tilde{V}_b = \left[ {v}^1_{b,1},\ldots, {v}^{N_{v,b}}_{b,1}, \ldots, {v}^{1}_{b,d}, \ldots, {v}^{N_{v,b}}_{b,d} \right]^T \text{ and }  \tilde{P}_b = \left[p_b^1, \ldots p_b^{N_{p,b}} \right]^T
\end{align*}
The $\tilde{V}_e$ and $\tilde{P}_e$ vectors are defined analogously.

Algebraically, we can re-write the PUFEM weak form in Eq.~(\ref{eq:PUFEM_Stokes_weak_form}) as:
\begin{equation}
	\left[\begin{array}{cccc}
		\bA_{bb} & \bA_{be} & \bB_{bb} & \bB_{be} \\
		\bA_{eb} & \bA_{ee} & \bB_{eb} & \bB_{ee} \\
		-\bB^T_{bb} & -\bB^T_{eb} & \mathbf{0} & \mathbf{0}\\
		-\bB^T_{be} & -\bB^T_{bb} & \mathbf{0} & \mathbf{0}
	\end{array}\right]
	\left[\begin{array}{c}
		\tilde{V}_b \\
		\tilde{V}_e \\
		\tilde{P}_b \\
		\tilde{P}_e
	\end{array}\right] =
	\left[\begin{array}{c}
	\bd_b \\
	\bd_e \\
	\mathbf{0}\\
	\mathbf{0}
	\end{array}\right]
	\label{eq:algeb_sys}
\end{equation}
Here the $\bA$ and $\bB$ blocks result from the Laplacian and incompressibility terms, respectively. Their general structure takes the form:
\begin{equation*}
	\bA_{\alpha\beta} = \left[ \begin{array}{ccc}
	\bA_{\alpha\beta} & \mathbf{0} & \mathbf{0}  \\
	\mathbf{0} & \ddots & \mathbf{0} \\
	\mathbf{0} & \mathbf{0} & \bA_{\alpha\beta}
	\end{array} \right]
	\text{ and }
	\bB_{\alpha\beta} = \left[ \begin{array}{c}
	\bB_{\alpha\beta}^1 \\
	\vdots \\
	\bB_{\alpha\beta}^d
	\end{array}  \right]
\end{equation*}
where $\alpha$ and $\beta$ are placeholders for any $b$ and $e$ permutation. The inner sub-blocks are defined as:
\begin{subequations}
	\begin{align}
          (\bA_{bb})_{mn} &= \int_{\Omega^h_b} \mu\nabla\left[(1-\psi^h)\hat{\phi}^m_v\right]\cdot\nabla\left[(1-\psi^h)\hat{\phi}^n_v\right]\hspace{1mm} d\Omega
          \label{eq:A_bb} \\
          (\bA_{be})_{mn} &=  \int_{\Omega^h_e} \mu\nabla \left[(1-\psi^h)\hat{\phi}^m_v\right]\cdot 	\nabla\left[\psi^h\hat{\zeta}^n_v\right]\hspace{1mm} d\Omega \\
          (\bA_{eb})_{mn} &= \int_{\Omega^h_e} \mu \nabla\left[\psi^h\hat{\zeta}_v^m\right]\cdot \nabla \left[ (1-\psi^h) 	\hat{\phi}^n_v \right]  \hspace{1mm} d\Omega \\
          (\bA_{ee})_{mn} &= \int_{\Omega^h_e} \mu  \nabla \left[\psi^h\hat{\zeta}^m_v\right]  \cdot \nabla \left[\psi^h \hat{\zeta}^n_v \right]  \hspace{1mm} d\Omega \\
          (\bB_{bb}^i)_{mn} &= -\int_{\Omega^h_b} \left[(1-\psi^h)\hat{\phi}^n_p\right]  \frac{\partial}{\partial x_i} \left[(1-\psi^h)\hat{\phi}^m_v  \right] \hspace{1mm} d\Omega \\
          (\bB_{be}^i)_{mn} &= -\int_{\Omega^h_e} \left[\psi^h\hat{\zeta}^n_p\right]\frac{\partial }{\partial x_i} 	\left[(1-\psi^h)\hat{\phi}^m_v\right] \hspace{1mm} d\Omega \\
          (\bB_{eb}^i)_{mn} &= -\int_{\Omega^h_e}  \left[(1-\psi^h)\hat{\phi}_p^n\right] \frac{\partial}{\partial x_i}   \left[\psi^h\hat{\zeta_v^m}\right]\hspace{1mm} d\Omega \\
          (\bB_{ee}^i)_{mn} &= -\int_{\Omega^h_e}  \left[\psi^h\hat{\zeta}_p^n\right] \frac{\partial}{\partial x_i}   \left[\psi^h\hat{\zeta_v^m}\right]\hspace{1mm} d\Omega
	\end{align}
	\label{eq:PUFEM_Stokes_blocks}
\end{subequations}
Vectors $\bd_g$ and $\bd_e$ are the right hand side vectors resulting
from imposed Dirichlet conditions. In order to apply the constraints
on the system matrix, we need to distinguish between the Dirichlet
condition and the constraints defined in
Section~\ref{sec:Redundant_DOF}.  While the former is trivial, in the
case of the latter, the original system matrix line is replaced with a
set of local basis functions evaluated at the constrained coordinate.
For example, if we wanted to constrain an arbitrary background velocity
degree of freedom, $v^m_{b,n}$, with the spatial coordinate $\bx_{b,n}$,
 the new line of the system of equation 
would be equivalent to:
\begin{equation}
	v^m_{b,n} - \sum_{i\in X_m} \hat{\zeta}^i_v(\bx_{b,n}) v^i_{b,n} = 0
\end{equation}
All other nodes, whether background or embedded, velocity or pressure,
are fixed analogously.We use $\bX_m$ to denote the set of nodes of 
embedded element in which
we find $\bx_{b,n}$.
We use $\bX_m$ to denote the set of nodes of embedded element in which
we find~$\bx_{b,n}$.
In the next section,  we expound the process involve in computing the
weighted weak form integral defined in ~(\ref{eq:PUFEM_Stokes_blocks}).
                
%%% Local Variables:
%%% mode: latex
%%% TeX-master: "Max_Article"
%%% End:

\subsection{Computing the PUFEM weak form}
\label{sec:alg_intersection}
\begin{figure}[!]
	\centering
	\includegraphics[trim={0cm 0cm 0cm 0cm},clip,page=2,width=\textwidth]{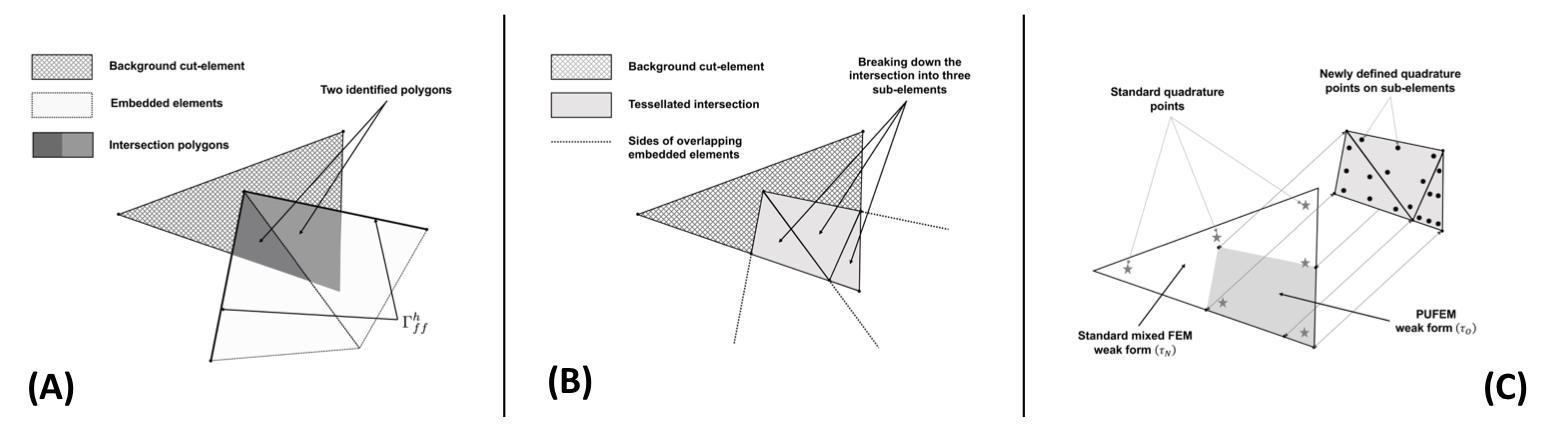}
	\caption{The intersection of one background element and two embedded elements at different stages of computing the PUFEM weak form. \textbf{(A)} Identifying the intersection polygons. \textbf{B} Creating a new tessellation for the overlap area. \textbf{(C)} Defining new set quadrature points for the sub-elements and re-evaluating the basis functions. } 
	\label{fig:intersection}
\end{figure}

We will start by illustrating this process with $\bA_{bb}$. The definition of this block, 
as defined in Eq.~(\ref{eq:A_bb}) can be rewritten as the sum of a classic weak form matrix,
 defined on the background mesh, and a mixed classic-PUFEM matrix defined on the overlap:
\begin{align}
	\left(\bA_{bb}\right)_{mn} & = \int_{\Omega^h_b} \mu \nabla \hat{\phi}^m_v : \nabla \hat{\phi}^n_v~d\Omega
	\nonumber \\
	 & + \int_{\Omega^h_e} - \mu \nabla \hat{\phi}^m_v : \nabla \hat{\phi}^n_v + 
	 \mu\nabla \left[\left(1-\psi^h \right)\hat{\phi}^m_v\right]:\nabla \left[\left(1-\psi^h \right)\hat{\phi}^n_v\right]
\end{align}
Thus we can separate the construction of $\bA_{bb}$ into two stages: 
first building a Laplacian block using classic FEM and 
second subtracting the classic weak form from the overlap area and replacing
it with the PUFEM version. While the former is trivial, the later is more challenging
as the fields and integration area are defined on non-matching meshes. In order to 
overcome this, we construct an intersection or tertiary mesh of sub-elements which allows us relate
the background and embedded meshes. Thus, at the start of the second stage,
for each background element we want to identify all potentially intersecting 
embedded elements. This can be
achieved optimally by adapting one of several approaches found in the
literature~\cite{massing2013efficient,mayer2009interface,wang2012computational}.
We then test the pairs of background and embedded candidates for
intersection and we identify the area of overlap, see
Fig.~\ref{fig:intersection}. If the area is non-zero, then the
resulting intersection will be a convex polygon with up to 6
sides. Each area is subdivided into up to 4 subelements using a
Delauney triangulation algorithm and these are stored in the tertiary
mesh. The quality of this new mesh is not relevant since it is used
for integration exclusively. To finalize the process, we loop over the
intersection mesh and we update the block by subtracting the overlap 
classic weak form and replacing it with the PUFEM one.

Both $\bA_{be}$ and $\bA_{eb}$ are defined exclusively on $\Omega_e$
and thus do not require the first stage introduced $\bA_{bb}$.
Since both are defined using fields from both meshes, the blocks are constructed
by looping over the intersection mesh.
In the case of $\bA_{ee}$, the block can built using the classic approach
due to matching topologies for the space and fields.

The $\bB$ blocks are generated following an analogous procedure.
%%%%
                
%%% Local Variables:
%%% mode: latex
%%% TeX-master: "Max_Article"
%%% End:

% \subsection{Classic FEM and PUFEM approaches to the ALE Navier-Stokes }
\section{A PUFEM approach to ALE Navier-Stokes and FSI problems }
\label{sec:PUFEM_4_ALE_and_FSI}
Building on the foundations of Section~\ref{sec:method}, in this section, 
we present the application of PUFEM in the case of more complex settings,
namely for ALE Navier-Stokes and FSI problems. Similarly to the Stokes flow
case, we begin with a short review of the classic ALE approach 
(Section~\ref{sec:Classical_ALE}), followed by a presentation of the PUFEM
version (Section~\ref{sec:PUFEM_ALE}), where the focus is on how the 
method is adapted to deal with having a transient overlap between the two
meshes. Finally, in
Section~\ref{sec:PUFEM_FSI} we outline a method of coupling the fluid
solver with a hyper-elastic solid problem.
	
\subsection{Classic FEM approach to the ALE Navier-Stokes problem}
\label{sec:Classical_ALE}
Let us consider a time dependent physical domain
$\Omega_f\subset\realset^d\times\left[0, T\right]$.
 At each time point $t\in[0,T]$, we use the
$\Omega_f(t)$ notation to refer to the current spatial configuration of the fluid domain
and let boundary $\Gamma(t)$ be its boundary. We also define the reference domain
$\Lambda_f\subset\realset^d$ which can be bijectively mapped to
$\Omega_{f}(t)$, for all $t\in\left[0,T\right]$. Hence,
$\mathcal{P}:\Lambda_f\times\left[0,T\right] \rightarrow \Omega_f$
denotes a mapping function used to relate the reference and spatial
domains. The non-conservative incompressible Navier-Stokes equation in
arbitrary Lagrangian-Eulerian (ALE) form can be written as
follows~\cite{bazilevs2013computational,nordsletten2010preconditioner}:
find $(\bv,p)$ such that
\begin{subequations}
  \label{eq:ALE}
  \begin{align}
    \rho\partial_t \bv + \rho(\bv - \hat{\bv} )\cdot\nabla\bv - \divergence \bsigma_f& = \mathbf{0} && \text{ in } \Omega_f(t), \\
    \divergence \bv & = 0 && \text{ in } \Omega_f(t), \\
    \bv & = \bv_D && \text{ on } \Gamma_{f}^D(t), \label{eq:Dir_cond_1}\\
    \bv & = \bv_S && \text{ on } \Gamma_{fs}^D(t), \label{eq:Dir_cond_2} \\
    \bsigma_f \cdot \bn & = \bt_N &&  \text{ on } \Gamma^N_{f}(t),  \\
    \bv(\cdot, 0)  & = \bv_0  && \text{ in } \Omega_{f}(0),
  \end{align}
\end{subequations}
for all $t\in\left[0,T\right]$. 
Compared to the Stokes flow in~(\ref{eq:Stokes_strong}), the
fluid model is augmented with a series of additional elements. The
stress tensor is replaced by the symmetric Cauchy stress tensor:
\begin{equation} \bsigma_f(\bv) = \mu\left(\nabla \bv +
\nabla\bv^T \right) - p\bI
\end{equation}
Inertial effects are introduced using the time derivative and
nonlinear advection terms. The $\partial_t(\cdot)$ operator is the
time derivative with respect to a fixed point in $\Lambda_f$. The ALE
advection term is used to account for the arbitrary motion of the
domain, where the arbitrary domain velocity field is defined as
$\hat{\bv} = \partial_t\mathcal{P}$~\cite{nordsletten2010preconditioner,bazilevs2013computational}. 
Finally, we continue to assume that $\bt_n = \mathbf{0}$.
	
Using BDF(2) discretization, the classic non-conservative FEM problem
for equation~(\ref{eq:ALE}) at a given time step $n$ can be written
as~\cite{nordsletten2010preconditioner}: find
$(\bv^h_n, p^h_n)\in\boldsymbol{\mathV}^h_{D}\times \mathW^h$ such
that
$\forall \hspace{1mm} (\bw^h,q^h)\in\boldsymbol{\mathV}^h_{0} \times
\mathW^h$ we have
\begin{align}
  \int_{\Omega_f^h(t_n)} \rho\left[\frac{3\bv^h_n - 4\bv_{n-1}^h + \bv_{n-2}^h }{2\Delta t} + \left(\bv^h_n-\hat{\bv}^h_n\right)\cdot\nabla\bv_n^h\right]\cdot\bw^h \hspace{1mm}d\Omega + & \nonumber \\
  \int_{\Omega_f^h(t_n)} \bsigma_f(\bv^h):\nabla\bw^h  + q^h\divergence\bv^h_{n} \hspace{1mm} d\Omega = & 0
\end{align}
Here, we consider the $\bv_{n-1}^h,\bv_{n-2}^h\in\boldsymbol{\mathV}^h$ 
to be the velocity fields at times $t_{n-1}$ and $t_{n-2}$, respectively, 
which have been transported via ALE mapping into the current spatial configuration. 

Due to the non-linearity of the system, we approximate the
solution using a Newton-Raphson (NR) based
algorithm~\cite{hessenthaler2017validation,shamanskii1967modification}. The
problem is solved semi-monolithically in that the mesh
velocity field, $\hat{\bv}^h_n$, is solved separately. If
$\bv_s$ is provided analytically or is obtained from a
weakly-coupled solid solver, then this process can be done at
the beginning of each time step. If the fluid and solid
solvers are coupled in a monolithic system, then the mesh
velocity is recomputed prior to each NR iteration in order to
account for the current estimate of the solid deformation.
        
%%% Local Variables:
%%% mode: latex
%%% TeX-master: "Max_Article"
%%% End:

\subsection{Computing mesh velocity}
\label{sec:comp_grid_vel}

The deformation of the mesh can be determined by an arbitrary problem
which is guided by the known deformations on portions of the
boundary. It is also well known that the choice of the problem can
significantly impact mesh quality over time. For this reason, we
propose to propagate the deformation of the boundary onto the entire
mesh by solving a solid mechanics problem based on the
nearly-incompressible neo-Hookean material model. In order to prevent
an excessive deterioration of the mesh we consider the material to be
heterogeneous as we allow each element to stiffen in a squared inverse
relationship with a metric of its relative element quality. A similar
idea has been successfully implemented in~\cite{spuhler20183d}.

Let us consider the case where we want to compute the mesh deformation
for time $n$. We choose $\Lambda_f$ to be the reference undeformed
domain. Thus the solid problem can be written as follows: find the
arbitrary domain deformation field $\hat{\bu}$ such that
\begin{subequations}
	\begin{align}
	\nabla_0 \cdot\bP_g  &= \mathbf{0}  && \text{in } \Lambda_f, \\
	\hat{\bu} & = \hat{\bu}_D && \text{on } \Gamma^D_{ref}, \\
	\bP_g\cdot\bN & = \mathbf{0} && \text{on } \Gamma^N_{ref}.  	
	\end{align}
\end{subequations}
This relates to the domain deformation velocity as
$\hat{\bv} = \partial_t \hat{\bu}$. Here, the $(0)$ subscript
indicates that the divergence operator is defined in the Lagrangian
coordinates and the first Piola-Kirchhoff tensor $(\bP_g)$ is given
as:
\begin{equation}
	\bP_g=  \mu_g\left[\frac{1}{J} \left(\bF -\frac{\bF:\bF}{2} \bF^{-T} \right) + \kappa J(J-1)\bF^{-T}\right]
	\label{eq:PK_nearly_incompressible}
\end{equation}
where $\bF = \nabla_0 \hat{\bu}$ is the deformation gradient and
$J = \det(\bF)$.
The heterogeneous stiffness parameter , $\mu_g$, is
defined as:
\begin{align*}
\mu_g(\tau) = \mu_{ref}\left( \frac{Q(\tau_0)}{Q(\tau)}\right)^2 \hspace{.5cm} \text{ where }  \hspace{.5cm} Q(\tau) = \frac{r}{R}
\end{align*}
Here $\mu_{ref}$ is the reference stiffness;
$Q:\mathcal{T}\rightarrow\realset_{+}$ is an element quality metric
defined as the ratio between the incircle ($r$) and circumcircle ($R$) of input
triangular element; $\tau_0$ and $\tau$ denote the original and
current shape of the element, respectively. As the quality of the
element deteriorates, $Q(\tau)$ decreases, making it stiffer and,
thus, less likely to further deform. $\kappa = 10$ is a penalty factor
on volumetric deformation. This choice was done heuristically and it
aims to penalize excessive decreases in element size.

In the discrete time setting, we chose to use the previous time step
configuration as the reference, undeformed domain. For the
normalization factor $Q(\tau_0)$, we use the element at time zero as
our reference. Given a discrete time point $t_n$, we use the
deformation field to compute the current space mapping function
follows:
\begin{equation}
	\mathcal{P}^h(\bX^h,t_n) = \mathcal{P}^h(\bX^h,t_{n-1}) + \hat{\bu}_n^h(\bX^h)
\end{equation}
Based on this, the mesh velocity is defined as $\hat{\bv}^h_n = \frac{\hat{\bu}_n^h}{\Delta t}$.

\subsection{PUFEM approach to the ALE Navier-Stokes problem}
\label{sec:PUFEM_ALE}

In Section~\ref{sec:PUFEM_Stokes} we discussed the setup of PUFEM for
flow around a rigid solid. Now we expand these original concepts in
order to treat the case where the two meshes are allowed to move
independently from each other.
	
A key requirement for the extension of PUFEM to ALE is the treatment
of the material time derivative.  Let $\Lambda_b^h$ and $\Lambda_e^h$
be the two reference domains for the background and embedded
components.
$\mathcal{P}_b:\Lambda_b^h\times\left[0,
  T\right]\rightarrow\Omega_b^h$ and
$\mathcal{P}_e:\Lambda_e^h\times\left[0,T\right]\rightarrow\Omega_e^h$
are the two arbitrary mapping functions, where $\Omega_b^h$,
$\Omega_e^h \in \realset^d\times[0,T]$ are the respective physical
domains. We also introduce the $\partial^b_t(\cdot)$ and
$\partial^e_t(\cdot)$ as the partial time derivatives with respect to
fixed points the global and embedded reference domains, respectively.
For ease, we consider the ALE expansion of the material time
derivative on an arbitrary semi-discrete (in space) scalar field
$f^h(t) \in \bV^h$ which can be written in PUFEM form as
$f^h = (1-\psi^h) f^h_b + \psi^h f^h_e$, where
$f_k^h(t)\in V^h_{k,*} $, $k \in \{e,b\}$.
Based on this, we can split the
time material derivative of $f$ into three terms:
\begin{equation*}
  \frac{D f^h}{D t} = \frac{D \left[(1-\psi^h)f_b^h + \psi^h f_e^h\right]}{D t} = (1-\psi^h)\frac{D f_b^h}{D t} + \psi^h \frac{D f_e^h}{D t} + (f_e^h - f_b^h)\frac{D \psi^h}{D t}
\end{equation*}
Furthermore, we choose $\Lambda_b^h$ to be the reference
domain of $f_b^h$ and $\Lambda_e^h$ for $f_e^h$ and~$\psi^h$.

Thus, we can continue to expand the first term into
its ALE form such that:
\begin{align*}
  \left.\frac{D f_b^h}{D t}\right|_{\bx} & = \left.\frac{\partial f_b^h}{\partial t}\right|_{\bx} + \bv^h\cdot\nabla f_b^h = \left( \partial^b_t f_b^h - \frac{\partial \mathcal{P}^h_b}{\partial t } \cdot \nabla f_b^h  \right)  + \bv^h\cdot\nabla f_b^h \\
                                         & = \partial^b_t f_b^h  + (\bv^h - \hat{\bv}_b^h) \cdot \nabla f_b^h
\end{align*}
Similarly, we also get:
\begin{equation*}
  \left.\frac{D f_e^h}{D t}\right|_{\bx} =  \partial_t^e f_e^h  + \left( \bv^h - \hat{\bv}_e^h  \right) \cdot \nabla f_b^h
  \hspace{.5cm}\text{and} \hspace{.5cm}
  \left. \frac{D\psi^h}{Dt} \right|_{\bx}  = \left(\bv^h - \hat{\bv}_e^h\right) \cdot\nabla \psi^h
\end{equation*}
In the last case, we assumed that $\partial_t^e \psi^h = 0$. The total
material time derivative can thus be expressed as:
\begin{align*}
  \frac{D f^h}{D t} & = (1-\psi^h) \partial_t^b f_b^h+ \psi^h  \partial_t^e f_e^h + \bv^h\cdot\nabla f^h \\
                    & - (1-\psi^h)\hat{\bv}_b^h\cdot\nabla f_b^h - \psi^h\hat{\bv}_e^h\cdot \nabla f_e^h  - (f_e^h - f_b^h)(\hat{\bv}_e^h \cdot\nabla \psi^h)
\end{align*}

Moving to the discrete form, let us define $\Omega^h_{b,n}$ and
$\Omega^h_{e,n}$, the global and embedded meshes, respectively, for a
given time step $n$.  Given that $\bX^v_b$, $\bX^v_e$, $\bX^p_b $ and $\bX^p_e$ change
with the alignment of the meshes with respect to each other at each
time step, let us define $\bV^h_{D,n}$, $\bV^h_{0,n}$ and $W^h_{n}$ as
the test and trial spaces at discrete time $t_n$. Thus the new discrete
weak form of the ALE Navier-Stokes problem can be written as follows:
find $(\bv^h_n, p^h_n)\in\bV^h_{D,n}\times W^h_n$ such that for any
$(\bw^h,q^h)\in\bV^h_{0,n}\times W^h_n$ we have that
\begin{align}
  \int_{\Omega^h_{b}(t_n)} \rho \left(\frac{3 \bv^h_{n}   -  4\bv^h_{n-1} +\bv^h_{n-2} }{2\Delta t }  + \bv^h_n \cdot \nabla \bv^h_n\right)\cdot \bw^h~d\Omega  + & \nonumber\\
  \int_{\Omega^h_{b}(t_n)} - \rho\left[ (1-\psi^h_n)(\hat{\bv}^h_{b,n}\cdot\nabla)\bv^h_{b,n} + \psi^h_n(\hat{\bv}^h_{e,n}\cdot \nabla)\bv^h_e \right] \cdot \bw^h \hspace{1mm} d\Omega + & \nonumber \\
  \int_{\Omega^h_{b}(t_n)} - \rho(\hat{\bv}^h_e\cdot\nabla\psi^h_n) (\bv^h_{e,n} - \bv^h_{b,n}) \cdot \bw^h   + \bsigma^h_{f,n}: \nabla \bw^h- q^h\divergence \bv^h_n~d\Omega = & 0 
  \label{eq:PUFEM_ALE_disc_weak_form}
\end{align}
where $\bv^h_n = (1-\psi^h)\bv^h_{b,n} + \psi^h\bv^h_{e,n}$. Both
partial derivatives $\partial^g_t(\cdot)$ and $\partial^e_t(\cdot)$
have been approximated using the BDF(2) scheme. In order to solve this
system of equations, we again employ the technique described
in~\cite{shamanskii1967modification}
and~\cite{hessenthaler2017validation}. The structure of the resulting
Jacobian matrix is largely based on that of system matrix described in
Eq.~(\ref{eq:algeb_sys}). The changes arise in the sub-block $\bA$
where in addition to the Laplacian we also have mass and advection
terms resulting from linearisation of the residual:
\begin{align}
  \left(\bA^{ij}_{bb}\right)_{MN} &= \int_{\Omega^h_{b,n}} \rho\delta_{ij}   \left[\frac{3}{2\Delta t} \psi^h_b\psi^h_b\hat{\phi}^M_v\hat{\phi}^N_v + \psi^h_b\hat{\phi}^M_v \left(\bv_{n}^h\cdot \nabla\right) \left(\psi^h_b\hat{\phi}^N_v\right) \right]~d\Omega \nonumber  \\
                                  & + \int_{\Omega^h_{b,n}} \rho\psi^h_b\psi^h_b \left[\hat{\phi}^M_v\hat{\phi}^N_v\frac{\partial\left(\bv^h_n\right)_i}{\partial x_j} - \delta_{ij}\hat{\phi}^M_v\left(\hat{\bv}^h_{b,n}\cdot\nabla\right)\hat{\phi}^h_v \right]~d\Omega \nonumber\\
                                  & - \int_{\Omega^h_{b,n}} \rho\delta_{ij}\hat{\phi}^M_v\hat{\phi}^N_v \left(\hat{\bv}^h_e\cdot\nabla\psi^h_b\right) - \mu\delta_{ij}\nabla\left(\psi^h_b\hat{\phi}^M_v\right):\nabla\left(\psi^h_b\hat{\phi}^N_v\right)~d\Omega \nonumber \\
                                  & + \int_{\Omega^h_{b,n}} \mu \frac{\partial \psi^h_b \hat{\phi}^M_v}{\partial x_j} \frac{\partial \psi^h_b \hat{\phi}^N_v}{\partial x_i}~d\Omega
\end{align}
The other blocks resulting from permutations of background/embedded
test and trial functions are obtained analogously.  Here, $i$ and $j$
correspond to the spatial orientation of test and trial DOF,
respectively. $M$ and $N$ represent the indexes of the mesh nodes. For
ease, $\psi^h_b$ is used as shorthand for $(1-\psi^h)$. The
application of Dirichlet boundary conditions and additional
constraints is largely the same as the one described in
Section~\ref{sec:PUFEM_Stokes}. The exception is that while the
Jacobian is re-used for multiple time iterations, the constraints such
as $\bv_b = \bx_e$ for $\bx \in \bX^v_b$ need to be re-adjusted based
on the new overlap configuration.

\subsection{Transient fixed nodes in ALE context}
\label{sec:trans_fixed_nodes}
Moving the background and embedded meshes independently from
each other introduces new challenges: having a different area of 
overlap in consecutive time steps and, by extension, having 
transient sets $\bX^v_b$ and $\bX^p_b$. While the latter is of less 
importance, the former can be significant factor of 
instability due to the need to use previous time step solutions in order
to approximate the velocity time derivative, as shown in 
Eq.~\ref{eq:PUFEM_ALE_disc_weak_form}. Thus, if a node is fixed at one
time $t_n$ and subsequently active at time $t_{n+2}$, our estimation of the
acceleration is dependent on what we chose to assign $t_n$. While, this 
concern has been partly addressed in Section~\ref{sec:Redundant_DOF},
it only covers the case of background nodes which are found under
the fluid area of the embedded mesh. Thus, in this case we can assign them
a meaningful value through interpolation. This leaves, however, the case
of background nodes found under the solid. Here there are two scenarios.
In the first case our area of enrichment is sufficiently thick and the 
time step is sufficiently small such that when a background node, after
it leaves solid, remains in the overlap area for at least two more time
steps. This would allow for sufficient time for the node to obtain 
meaningful data in order to compute the derivative. However, this 
would be difficult to control, particularly in FSI. 
An alternative solution that we propose is to constrain the value of 
background nodes in the solid to be equal to the interpolated 
solid velocity field, which is generally continuous to that of the fluid flow.
If this field is provided or is computed as part of the FSI solution, then
this new task is trivial. If however, only the surface motion is provided,
one can potentially create a field by solving an artificial problem, such
as a diffusion one. It should be noted though that this last approach is only 
theoretical and was never implemented for this work.

%%% Local Variables:
%%% mode: latex
%%% TeX-master: "Max_Article"
%%% End:

\subsection{PUFEM for FSI}
\label{sec:PUFEM_FSI}

To illustrate the efficacy of the PUFEM approach for FSI, in this
section we elaborate on a monolithic approach to coupling the ALE
Navier-Stokes solver described in Section~\ref{sec:PUFEM_ALE} with a
generic quasi-static non-linear solid. Let $\Omega_{s,0}$ and
$\Omega_{s,t}\subset\realset^d$ be our reference and deformed (at time
$t$) solid domains.  Thus, the strong form of the FSI and ALE mesh
problems can be written as: find $(\bv_f, p_f, \bu_s,p_s,\hat{\bu}_e)$
such that
	\begin{align}
		% FLUID PROBLEM
		\begin{aligned}
		\rho\partial_t \bv_f+ \rho(\bv_f - \hat{\bv} )\cdot\nabla\bv_f - \divergence \bsigma_f & \\
		\divergence \bv_f & \\
		\bv_f& \\
		\bsigma_f \cdot \bn & \\
		\bv_f(\cdot, 0)  & 
		\end{aligned} 
		&
		\begin{aligned}
		& = \mathbf{0} \\
		& = 0 \\
		& = \bv_{f,D} \\
		& = \bt_{f,N} \\
		& =  \bv_0   
		\end{aligned}
		&&
		\begin{aligned}
		& \text{in } \Omega_f, \\
		& \text{in } \Omega_f, \\
		& \text{on } \Gamma_f^D \\
		&  \text{on } \Gamma^N_f, \\
		& \text{in } \Omega_{f,0},
		\end{aligned} \label{eq:FSI_fluid_problem}\\
		\hline
		% SOLID PROBLEM
		\begin{aligned}
		\nabla_0 \cdot \bP_s(\bu_s,p_s)  & \\
		J -1 & \\
		\bu_s & \\
		\bP_s(\bu_s,p_s) \cdot \bN &
		\end{aligned}
		&
		\begin{aligned}
		 & = \mathbf{0} \\
		 & = 0 \\
		 & = \bu_{s,D} \\
		 & = \bt_{s,N}
		\end{aligned}
		&&
		\begin{aligned}
		& \text{in } \Omega_{s,0} \\
		& \text{in } \Omega_{s,0} \\
		& \text{on } \Gamma^D_{s,0} \\
		& \text{on } \Gamma^N_{s,0}
		\end{aligned}
		\label{eq:FSI_solid_problem}
		\\
		\hline
		% FLUID-SOLID COUPLING
		\begin{aligned}
		\bsigma_f \cdot \bn +J^{-1} \bP_s(\bu_s,p_s) \cdot \bN & \\
		\bv_f & \\
		\end{aligned}
		&
		\begin{aligned}
		& = \mathbf{0} \\
		& = \dot{\bu}_s \\
		\end{aligned}
		&&
		\begin{aligned}
		& \text{on}~\Gamma_\mathit{fs} \\
		& \text{on}~\Gamma_\mathit{fs} \\
		\end{aligned} 
		\label{eq:FSI_coupling_conditions}
		 \\
		\hline
		% The ALE grid problem
		\begin{aligned}
		\nabla_0\cdot \bP_g(\hat{\bu}_e) & \\
		\hat{\bu}_e  & \\
		\bP_g(\hat{\bu}_e)\cdot \bN & 
		\end{aligned} 
		&
		\begin{aligned}
		& = \mathbf{0} \\
		& = \bu_f \\
		& = \mathbf{0}
		\end{aligned}
		&&
		\begin{aligned}
		& \text{in } \Omega_{e,0} \\
		& \text{on } \Gamma_{\mathit{fs},0}  \\
		& \text{on } \Gamma_{\mathit{ff},0} 
		\end{aligned}
		\label{eq:FSI_ALE_grid_problem}
	\end{align}
	The FSI problem can be divided into four parts: the fluid problem in~(\ref{eq:FSI_fluid_problem}), the quasi-static solid problem in~(\ref{eq:FSI_solid_problem}), the coupling conditions in~(\ref{eq:FSI_coupling_conditions}) and the ALE mesh problem in~(\ref{eq:FSI_ALE_grid_problem}).	For illustrative purposes, we chose to use the neo-Hookean model for the solid, resulting in the following form of the first Piola-Kirchhoff stress tensor:
	\begin{equation}
		\bP_s = \frac{\mu_s}{J_s} \left[ \bF_s - \frac{\bF_s : \bF_s}{2} \bF_s^{-T} \right] - p_s J_s \bF_s^{-T} 
	\end{equation}
	where $\bF_s = \nabla_0 \bu_s$ and $J_s = det(\bF_s)$.
	Note, we chose to simplify the solid problem by considering the loading process to be quasi-static. The constitutive law used for the arbitrary mesh deformation problem has been previously described in Section~\ref{sec:comp_grid_vel}. The PUFEM discrete setting is elaborated on in the next section.
	
	\subsubsection{Discrete weak form}
		The main difference compared to the ALE Navier-Stokes set-up described in Section~\ref{sec:PUFEM_ALE}  is that now we also need to solve the deformation/translation of the solid, rather than it being given.
		In order to take advantage of the definition of the PUFEM spaces in~(\ref{eq:PUFEM_Stokes_weak_form}),~(\ref{eq:local_spaces}) and~(\ref{eq:local_spaces_w_bc}) which extend into both $\Omega_f^h$ and $\Omega_s^h$, for each time step we compute the solid's velocity $\bv^h_n$, rather than its displacement. The deformation is obtained using the backward Euler scheme ($\bu^h_{s,n} = \bu^h_{s,n-1} + \bv^h_{s,n} \Delta t$). Thus, the problem coupling is ensured by the fact that the fluid and solid velocity DOF are identical on $\Gamma_{fs}$. 
		
		The fully discrete weak form  can be written as: find $(\bv^h_{n}, p_{n}^h) \in\bV^h_{D,n}\times W^h_n$ such that $\forall \hspace{1mm} (\bw^h, q^h)\in \bV^h_{0,n}  \times W^h_n$
		\begin{align}
		    \int_{\Omega^h_{f,n}} \rho\left(\frac{3\bv^h_{n} - 4\bv^h_{n-1} +\bv^h_{n-2} }{2\Delta t} + \bv^h_{n}\cdot \nabla \bv^h_{n} \right)   \cdot \bw^h~d\Omega & + \nonumber \\
			\int_{\Omega^h_{f,n}} -\rho \left[\left(1 - \psi^h_n\right)\hat{\bv}_{b,n}^h \cdot \nabla\bv^h_{b,n}  +\psi^h_n\hat{\bv}^h_{e,n} \cdot \nabla\bv^h_{e,n} \right] \cdot \bw^h \hspace{1mm} d\Omega & + \nonumber\\
			\int_{\Omega^h_{f,n}} -  \rho(\hat{\bv}^h_e \cdot \nabla\psi^h_n)  (\bv^h_{e,n} - \bv^h_{b,n}) \cdot \bw^h \hspace{1mm} d\Omega & + \nonumber \\
			\int_{\Omega^h_{f,n}}  \bsigma_{f,n}^h : \nabla\bw^h_{n}  + q^h \divergence \bv^h_{n} \hspace{1mm} d\Omega & + \nonumber \\
			\int_{\Omega^h_{s,0}} \bP_s : \nabla_{0} \bw^h + q^h(J_s - 1) 
			\hspace{1mm} d\Omega & = 0 
		\end{align}
		The structure of the system of equations which has to be solved during each Newton-Raphson remains largely the same. Note, the pressure is allowed to be discontinuous across $\Gamma_\mathit{fs}$

\subsection{Implementation}
In order to examine the performance of PUFEM against the classic mixed
FEM technique, we implemented both approaches in the MATLAB 2017b
language. The meshes used for the test were produced using the MESH2D
package~\cite{engwirda2005unstructured,engwirda2014locally}. All the
simulation where run on MacBook pro 2017 running on macOS Sierra, with
an Intel Core i7-7820HQ processors and 16 GB of 2133 MHz LPDDR3 RAM.

%%% Local Variables:
%%% mode: latex
%%% TeX-master: "Max_Article"
%%% End:

\section{Numerical Results}
\label{sec:results}
\subsection{Stokes flow convergence test}
\label{sec:res_stokes}

\begin{figure}
	\centering
	\includegraphics[trim={0 0cm 0 0},clip,width=.5\textwidth]{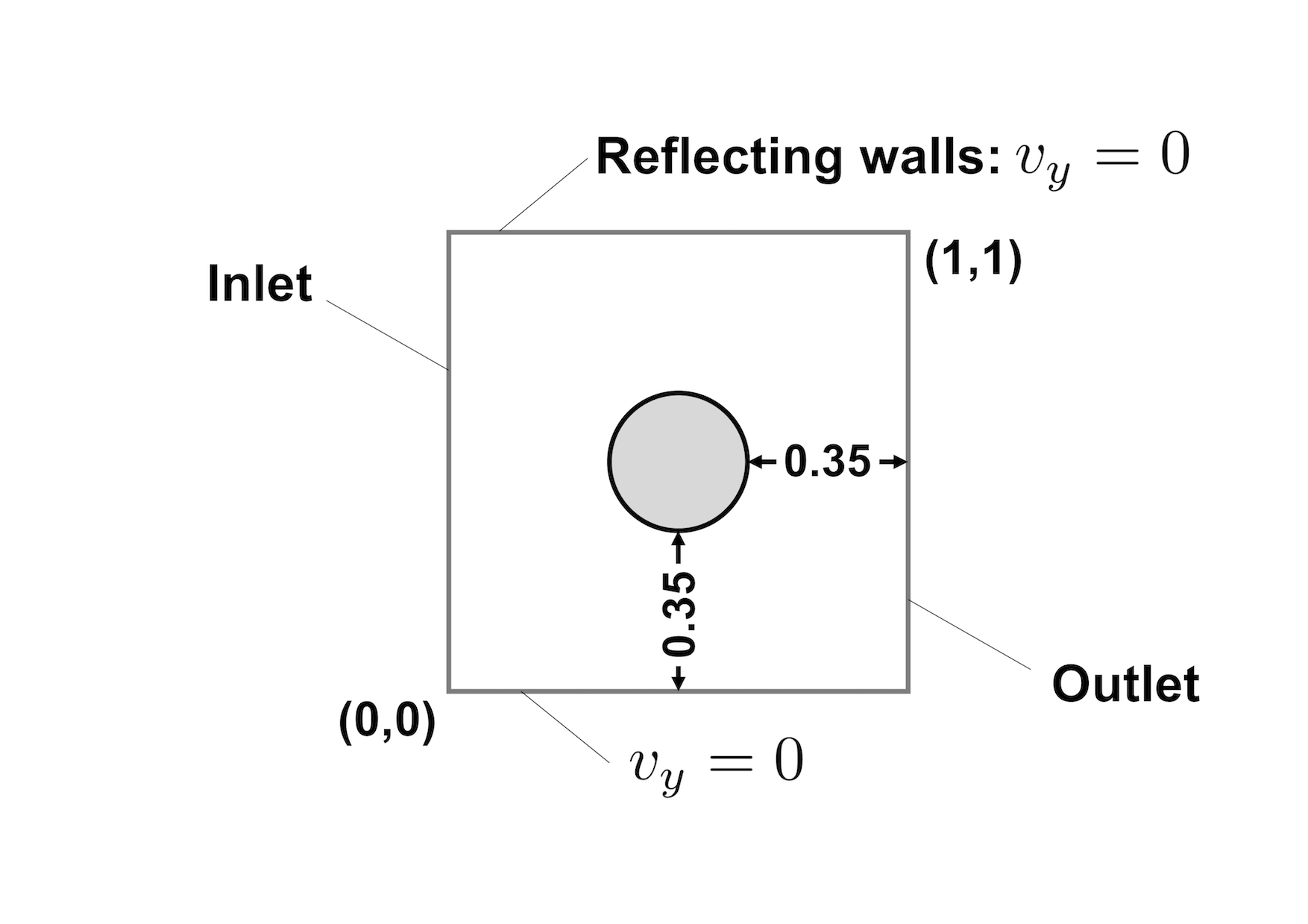}
	\caption{Domain used in the steady-state Stokes and Navier-Stokes problems.}
	\label{fig:stokes_domain}
\end{figure}

In order to illustrate the comparable accuracy of PUFEM to that of
standard boundary fitted FEM, we consider a basic steady-state
incompressible Stokes flow problem, where we compute the errors in the
velocity and pressure fields on series of meshes. More specifically, we
consider a total domain $\Omega_f\cup\Omega_s=\left[0,1\right]^2$
where a rigid cylindrical obstacle (with a radius of 0.15) is placed in the centre, see
Fig.~\ref{fig:stokes_domain}. The inflow (left side) is constrained using
the Dirichlet condition, $\bv_{inflow} = [1;0]^T$, which is constant 
throughout the boundary patch. On the side walls, we apply
a reflection Dirichlet condition on the velocity field, such that $v_y = 0$, 
and on the outflow we 
impose a zero traction condition. Fluid viscosity was set to 1.0.
The purpose of this arrangement is to concentrate
the boundary layer flow to the vicinity of cylinder, which coincides
with the area enriched by the embedded mesh.

To ease comparison, we try to use meshes with small variations in
element size. For
both approaches, we consider five refinement levels, with $h$ ranging
from approximately $0.1$ to $0.00625$. In the case of PUFEM we paired
the background and embedded meshes such that $h_g\approx h_e$. 
The latter mesh is composed of two regions: the cylinder, which is only
used to define the solid surface, and the fluid enrichment area, which forms
a ring with a thickness 0.161. We shall refer to this grid as M1. More details about 
their statistics and that of the classic mesh set can be found in Table~\ref{tab:mesh_stat}.

\begin{figure}[!]
		\centering
		\includegraphics[width=\textwidth]{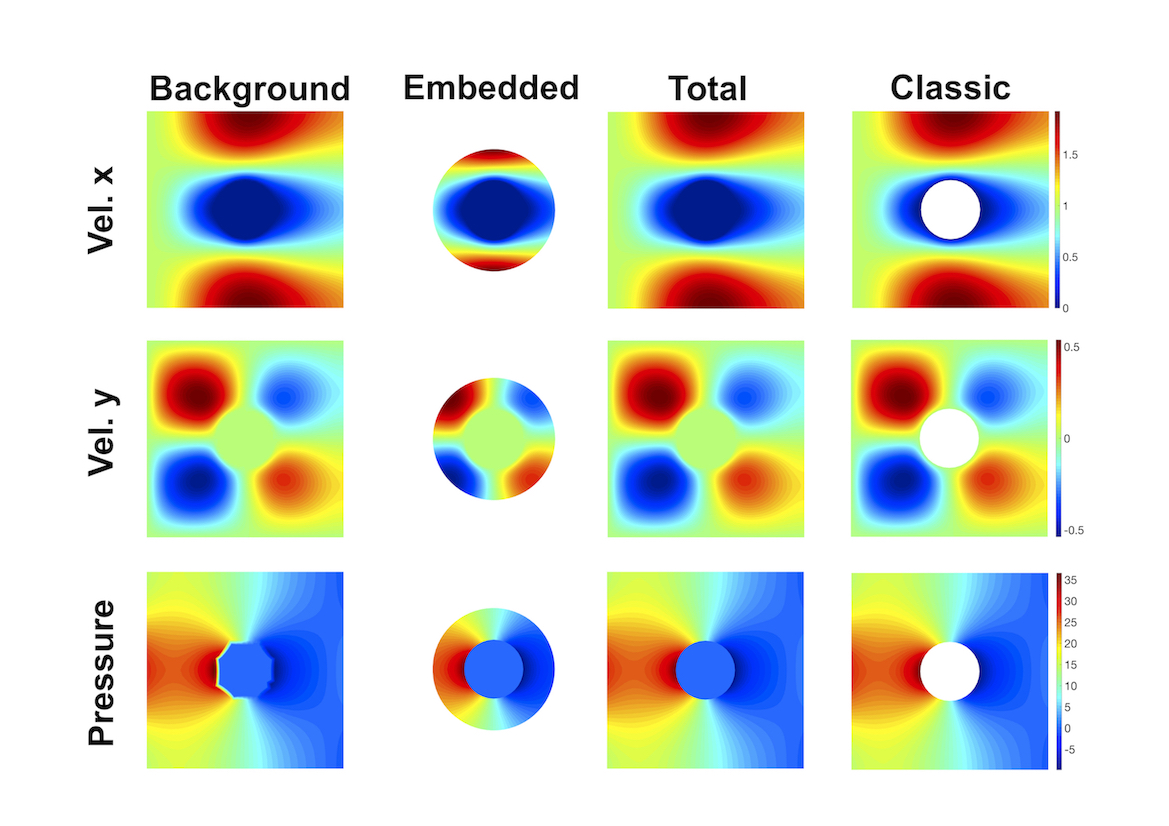}
		\caption{Example of solution fields for the Stokes problem obtained 
			using the PUFEM and classic approaches. The PUFEM figures include the
			background and embedded components, as well as the total 
			(resulting from the weighted sum).}
		\label{fig:Stokes_sol}
\end{figure}

The resulting velocity and pressure fields and their subcomponents can
be seen in Fig.~\ref{fig:Stokes_sol}. The PUFEM weighted sum result and
the classic solution appear to be very similar. Additionally, there are no 
obvious non-physical behaviours, such as field jumps,  which might have been
generated by the coupling.

\begin{figure}[!]
	\centering
	\includegraphics[trim={0cm 7cm 0 10cm},clip,width=\textwidth]{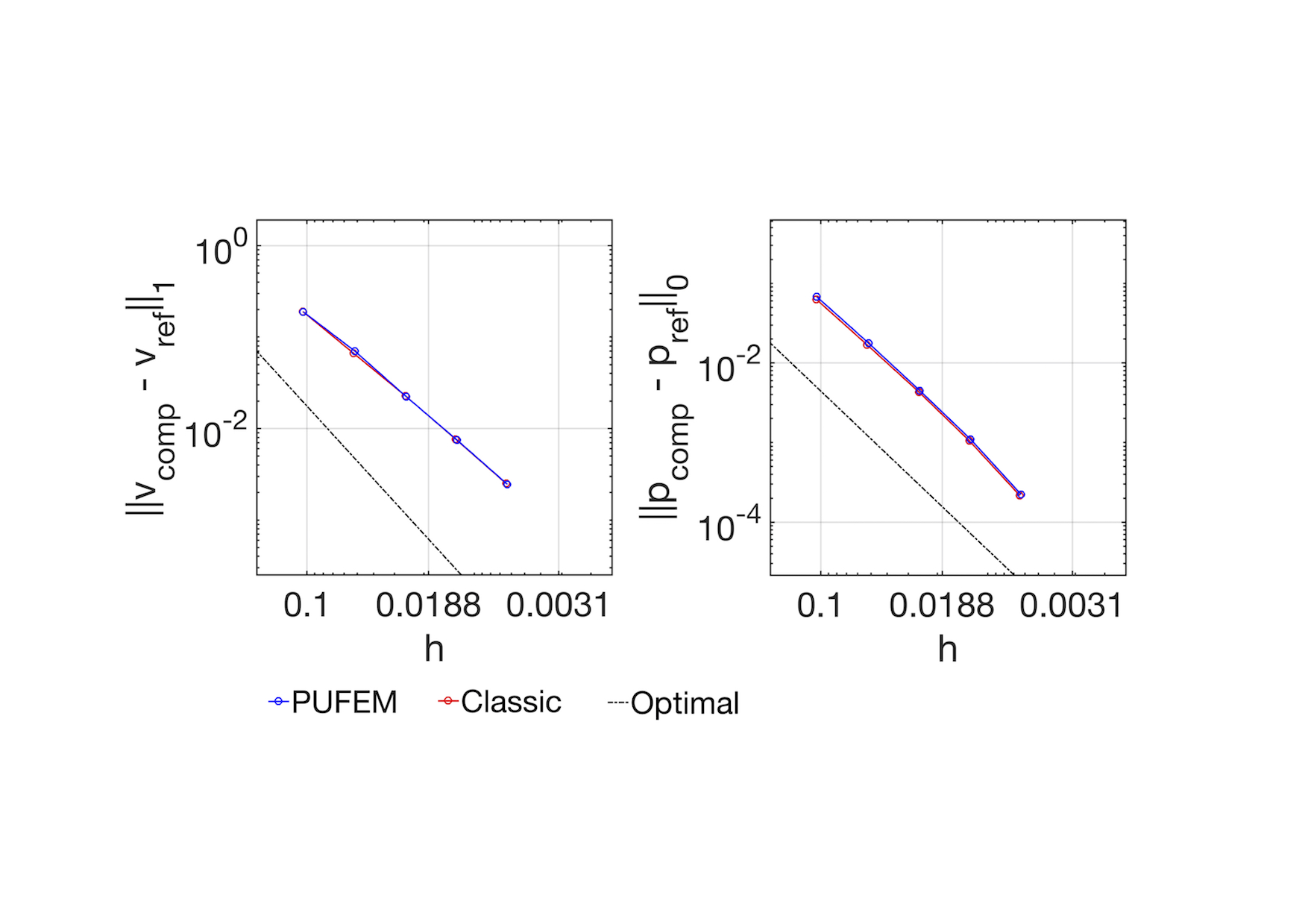}
	\caption{Plot showing the errors in the velocity and pressure fields for the Stokes flow problem obtained with the classical and PUFEM approaches. The dotted line indicates the optimal convergence rate.}
	\label{fig:res_stokes_convergence}
\end{figure}

The results of the convergence test are shown in
Fig. ~\ref{fig:res_stokes_convergence}.
Note, that we did not 
rely on an analytical reference solution to compute the error. In
order to obtain a reference, the flow field was computed using
the classic approach and a very fine, boundary fitted grid with an element
size of $h = 0.0016$.
We observe that velocity and
pressure fields seem to converge in the case of both PUFEM and
boundary fitted approach, with rates which closely follow the 
the \textit{a priori} estimate in Eq.~\ref{eq:ineq5}. We observe
that the velocity produces slightly suboptimal results, however,
this is likely due to both projection errors and non-conformity
between solutions. We also notice that the difference in errors
across the approaches are very small. Given the mesh size similarity
among the methods, PUFEM was not expected to exceed 
the performance of the classical approach. Thus, the accuracy of the new
method is ultimately capped by the local approximation capability of the
background and embedded grids.  Furthermore, the coupling, 
representing the main potential error source, 
appears to have little deteriorating impact.

\subsection{Steady-state Navier-Stokes convergence test}

To examine PUFEM performance for steady-state Navier-Stokes, we
consider the problem outlined in Fig~\ref{fig:stokes_domain} for
Reynolds numbers of 30 and 100. We keep using the $[0,1]^2$ domain with
$R =0.15$ cylindrical obstacle in the middle. The fluid parameters are set to:
$\rho = 1.0$ and $\mu = 0.01$. The boundary conditions are kept largely the 
same with the exception of the inflow velocity. Thus, the $x$ direction component
is set to $1.0$ or $10/3$, depending on whether we want to compute the
$Re$ 30 or 100 cases. Again, our errors estimates are based on flow results
computed using the classic approach and a fine grid with $h=0.0016$. 
This test allows us to look at any
potential deterioration in the accuracy of PUFEM due to introduction
of non-linearity. Moreover, by increasing the Reynolds number we
reduce the size of the boundary layer, which enables us to better
assess the benefit of using the embedded mesh to enrich the total
solution space. For PUFEM, we introduce a second mesh set,
denoted M2, where $h_b\approx 2h_e$. See Table~\ref{tab:mesh_stat}
for the corresponding mesh statistics.

\begin{figure}[!]
	\centering
	\includegraphics[width=\textwidth]{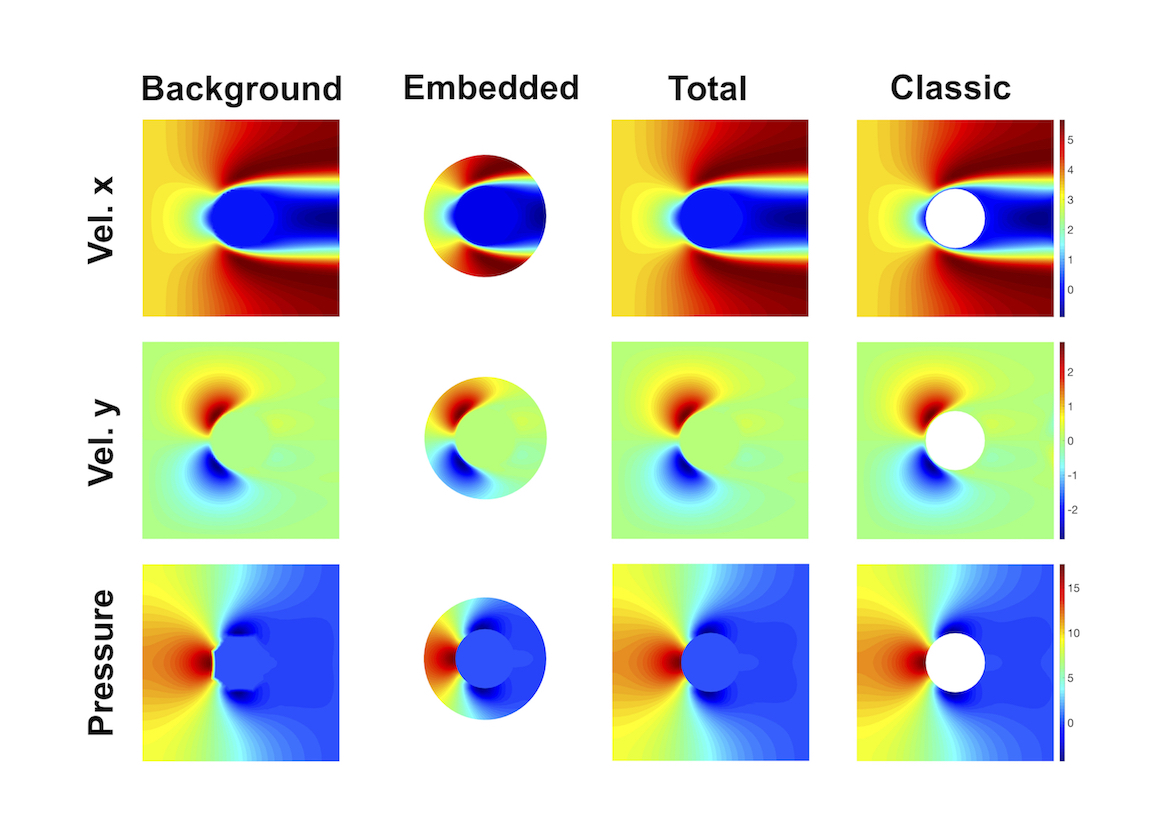}
	\caption{Example of solution fields for the Navier Stokes problem at $Re$ 100 obtained 
		using the PUFEM and classic approaches. The PUFEM figures include the
		background and embedded components, as well as the total 
		(resulting from the weighted sum).}
	\label{fig:NS_ss_sol}
\end{figure}

A set of sample results for the velocity and pressure at $Re$ 100
can be seen in Fig.~\ref{fig:NS_ss_sol}. Despite the increase in flow
complexity, the quality of the PUFEM solution remains comparable to
that of the classic result.

\begin{figure}[!]
	\centering
	\includegraphics[trim={0cm 5cm 0 10cm},clip,width=\textwidth]{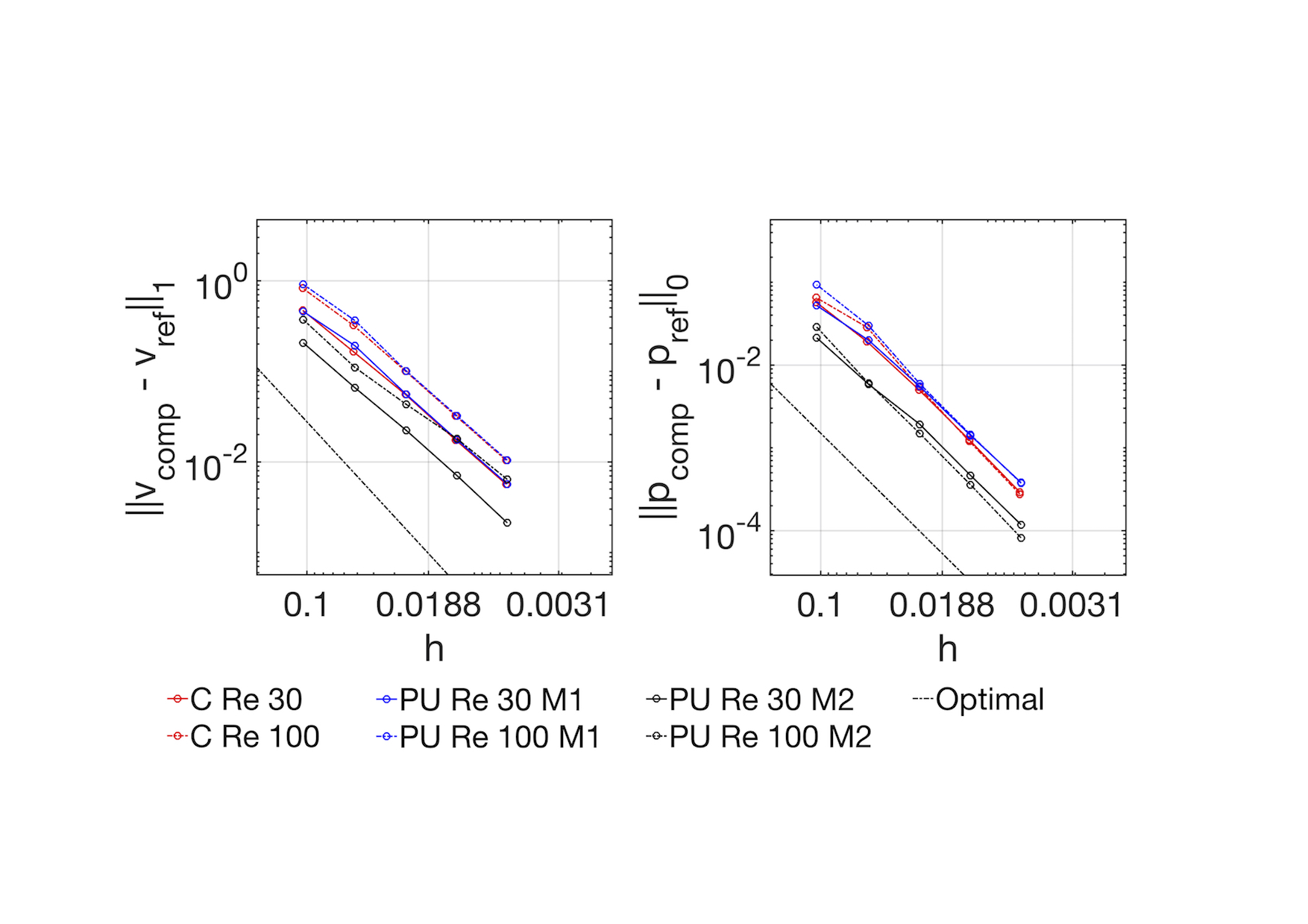}
	\caption{The error plots for steady state incompressible Navier-Stokes. Both classic (C) and PUFEM (PU) methods are run at Reynolds number of 30 and 100.}
	\label{fig:res_NS_convergence}
\end{figure}

The error plots for the different cases
are found in Fig.~\ref{fig:res_NS_convergence}.  A first observation is that
the convergence behaviour for $Re$ 30 and 100
 remains relatively unchanged from that observed in Stokes flow
 problem.
Furthermore, the differences in accuracy between classic and 
PUFEM (M1) approaches remain small. As previously noted in the 
Stokes case, when using similar spatial discretization, we do not
expect the new approach present a significant advantage. It continues 
to perform well in these circumstances and is seemingly unaffected by
the rise in $Re$. Switching to M2, we see an overall decrease in error.
This suggest that PUFEM can potentially leverage local solution enrichment
to improve overall accuracy. While a similar behaviour can be replicated in 
the case of boundary fitted grids using adaptive meshing, 
the main strength of the new method resides in 
its ability to better withstand large deformations.

\subsection{Sch\"{a}fer-Turek benchmark}
\label{ssec:schaf-turek-benchm}
\begin{figure}[!]
	\centering
	\includegraphics[trim={0cm 1cm 0cm  1cm},clip,width=\textwidth]{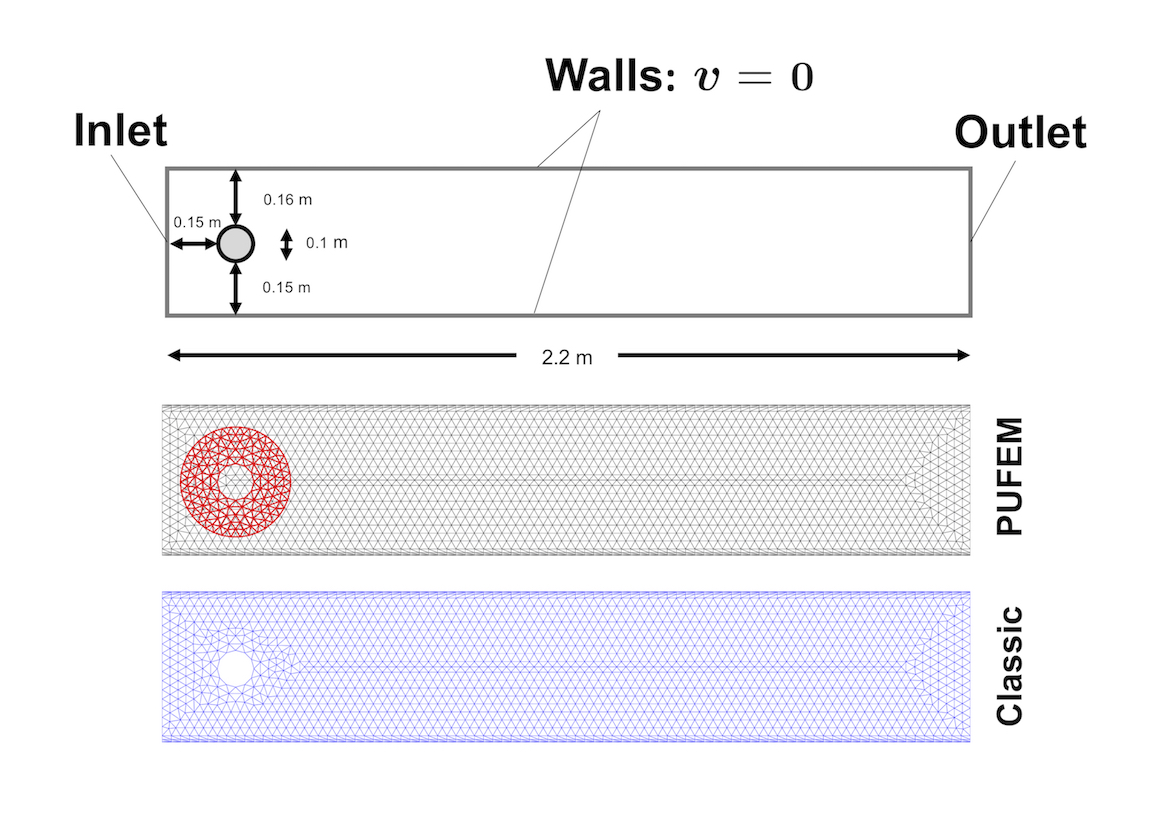}
	\caption{The domain of the Sch\"afer-Turek benchmark problem with boundary labels as well as example meshes
          used in PUFEM and classic approaches.}
	\label{fig:Turek_domain_meshes}
\end{figure}

In order to evaluate the stability of the PUFEM method in the case of
time dependent problems and also to compare it to other published flow
results, we considered the Sch\"{a}fer-Turek
benchmark~\cite{schafer1996benchmark}, case 2D-2 with $Re=100$ 
(see Fig.~\ref{fig:Turek_domain_meshes}). 

The inflow constraint is imposed on the velocity using the following
Dirichlet condition:
\begin{equation*}
	\bv = \left[ \begin{array}{c}
	\frac{4Uy(W-y) }{W^2} \min(t,1) \\ 0
	\end{array}\right]
\end{equation*}
After linearly ramping up, the inflow profile remains constant and we run
the problem for $10$ $s$ in simulation time, enough to initiate the
characteristic shedding and also to reach steady state oscillations.
Here $U = 1.5~m/s$ is the reference flow speed and $W = 0.41~m$ 
is the width of the tube. 
We also impose a no slip condition on the side walls and zero traction
on the outflow. In order to achieve the correct Reynolds number,
density ($\rho$) and viscosity ($\mu$) are 
set to $1~kg/m^3$ and $0.001~Pa\cdot s$, respectively.

In total, six solutions are produced using PUFEM based on two mesh
sets (PU1 and PU2) and three time steps (i.e. 0.01, 0.002 and 0.001 $s$). 
Each mesh set consists of a regular background grid and 
a boundary fitted embedded grid. The fluid region of the latter forms 
a ring with a thickness of 0.1 $m$ around the cylinder, 
see Fig.~\ref{fig:Turek_domain_meshes}.
For simplicity, each set uses a constant $h$ for both background and
embedded grids, with the value set to 0.025 and 0.0125 $cm$ in
the case of PU1 and PU2, respectively.  Additional mesh statistics
are found in Table~\ref{tab:mesh_turek}. To aid the comparison, we have 
also produced a set of flow results using the classical approach using
a quasi-regular grid with $h=0.025$ and a time step of 0.01~$s$.

\begin{figure}[!]
	\centering
	\includegraphics[trim={0cm 20cm 0cm 0cm},clip,width=\textwidth]{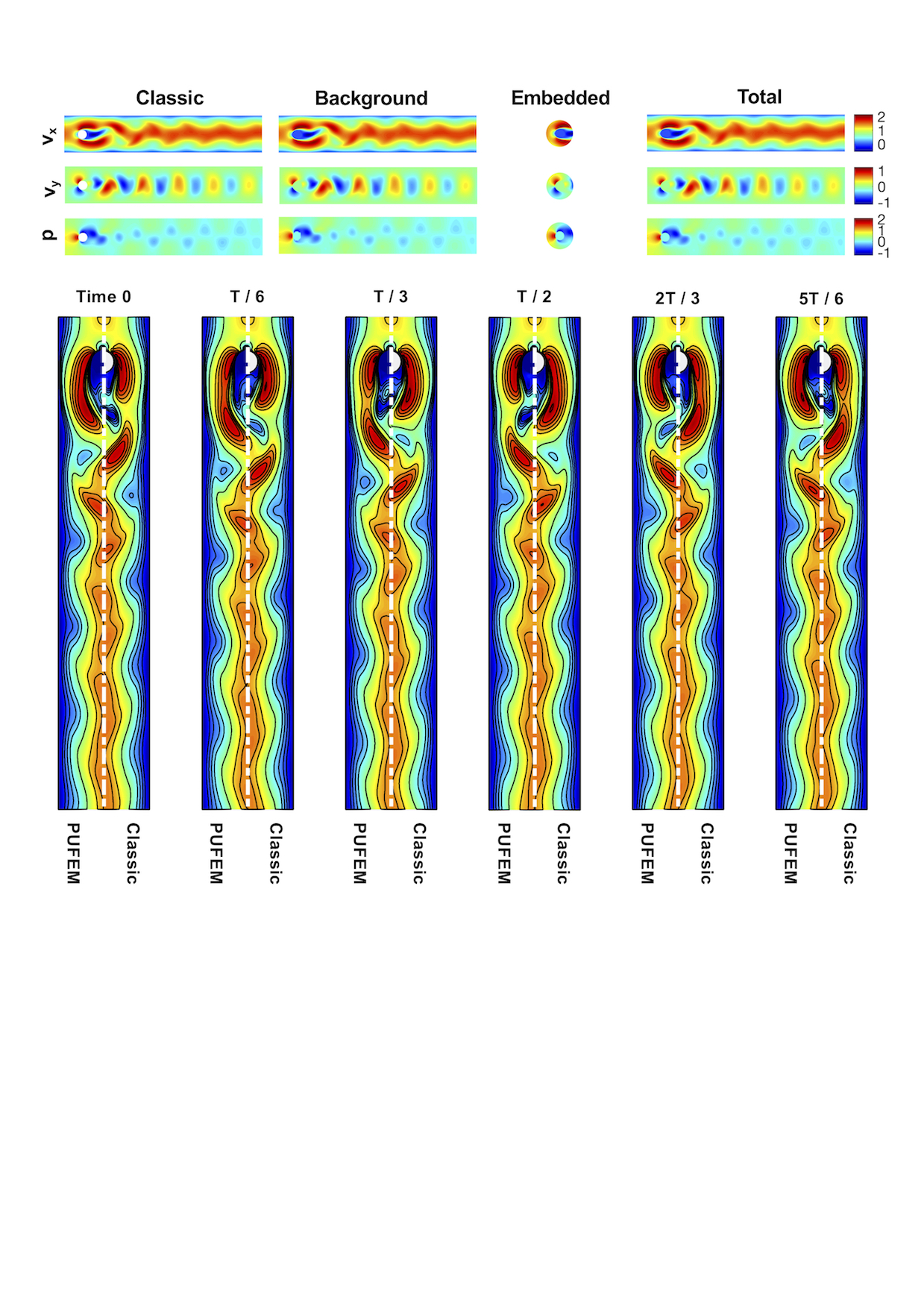}
	\caption{Compiled results for the Turek benchmark. (Top) A breakdown of the PUFEM solution at peak $c_l$ (9.57 $s$ in simulation time) into background, embedded and total fields for each component of the solution. The classic result is introduced for comparison. (Bottom) Velocity magnitude over a cycle in both PUFEM and classical simulations. Contour lines where added to aid comparison.}
	\label{fig:Turek_benchmark}
\end{figure}

Fig.~\ref{fig:Turek_benchmark} shows an example set of flow results
(i.e. velocity components and pressure) computed using classic and
PUFEM approaches. In the top part, we have a snapshot of the flow
at one of the time points of maximum lift (i.e. 9.57 $s$ in simulation time
for PUFEM). The results are split into subcomponents
corresponding to the 
boundary fitted mesh, background, embedded meshes
and the weighted sum.
In the second half of the figure, we include six 
velocity magnitude snapshots at the different stages of the shedding
process. Each frame is split in two, with the right side showing the 
PUFEM field and the left displaying the classic, showing
the wake patterns are in good agreement. Furthermore, the transition
between the embedded and background grids retains its smoothness,
despite the transient behaviour of the solution.

From a quantitative perspective, we also computed the coefficients of
drag ($c_d$), lift ($c_l$), Strouhal number($St$) and pressure drop
across the cylinder ($\Delta p$), defined as:
\begin{align*}
	c_d & = \frac{2 F_d}{\rho \bar{U}^2 D}, \hspace{5mm} 
	F_d = \int_{\Gamma_{\mathit{fs}}}  \mu \frac{\partial v_t}{\partial n} n_y -pn_x~d\Gamma \\
	c_l   & = \frac{2 F_l}{\rho \bar{U}^2 D} , \hspace{5mm}
	F_l =  - \int_{\Gamma_{\mathit{fs}}} \mu \frac{\partial v_t}{\partial n} n_x + pn_y~d\Gamma \\
	St  & = \frac{fD}{\bar{U}} ,
\end{align*}
where $D$ is the diameter and $v_t$ is the tangential fluid velocity.
These values have been compiled in Table~\ref{tab:coeff_turek},
which contains all the PUFEM and classic results, as well as 
those from literature. For similar space and time discretization,
i.e. entries two and three of the table, the new and classic 
approaches provide similar estimations of the with errors 
of: $0.03 \%$ for $c_d$, $0.22 \%$ for $c_l$, $3\%$ for
$St$, and $0.16\%$ for $\Delta p$. By comparing to the literature
values~\cite{schafer1996benchmark}, we see that the PUFEM estimations
of $St$ and $\Delta p$ are converging inside the recommended
interval. However, $c_d$ and $c_l$ are generally underestimated, with
minimum errors of $1.18\%$ and $1.41\%$, respectively, achieved using
the PU2 mesh set and 5000 time steps. Generally, though, the 
trend suggests that all four parameters would converge 
provided sufficient spatial and temporal discretization. 

 Given the initial stage of the development, it has not been our aim to
 produce an optimal implementation of PUFEM. However, this has 
 represented a significant 
obstacle to running very large problems. For example, the total
solving time for the classic and new approaches using the mesh
sets FEM1 and PU1, respectively, and a time step of 0.01~$s$ 
were approx. 1.9 and 
2.9 hours, respectively. In this case, where the mesh overlap is fixed, 
the main factor responsible PUFEM's lag appears to 
be the residual's assembly, with an average running time 0.42~$s$, compared
to 0.097~$s$ for the classic. While not addressed in this work, there are
multiple avenues which could be used to reduce the this disparity:
restricting the interface mesh to the area where $\psi$ is strictly between
0 and 1, contracting this area to a thinner strip close to $\Gamma_\mathit{ff}^h$ and
optimizing the quadrature rule being used, to name a few.

\subsection{ALE: An oscillating rigid cylinder}
\label{ssec:ale-oscill-cyl}
Moving towards moving domain applications, we define a simple ALE
benchmark problem. Using a similar to setup to that of the Turek
benchmark, we now place the cylinder in the middle of the domain, and
we oscillate it along the long axis of the tube, while we allow for
free outflow at the left and right ends. The cylinder displacement 
as a function of time is defined as:
\begin{equation}
d (t)= \left\{
\begin{array}{cc}
A\frac{\sin(\omega t)}{\omega} - At \text{ } \cos(\omega t), & t \leq 1 \\
-A\text{ } \cos(\omega t) ,& t > 1
\end{array}
\right.
\end{equation}
where $A$ is the maximum amplitude and $\omega$ is the frequency. We
chose to set $A~=~0.2$ m and $\omega = 2\pi \hspace{1mm}s^{-1}$ ,
resulting in a maximum velocity of approx $1.25 \hspace{1mm} ms^{-1}$
and a $Re\approx 100$. We are interested in the first 2 seconds of the
simulations. The goal of this test is to verify the stability of the
PUFEM approach in the context of a transient overlap area.  For this
reason, the amplitude of the displacement was chosen large enough to
cause multiple sharp changes in $\Omega_{cut}$.

For the spatial discretization, we define a new ALE-PU1 mesh set to be
consistent with the new geometry (i.e. in accordance to the
repositioned cylinder) and with identical mesh statistics as PU1 (i.e. due
to recycled connectivity matrices). For
the classical approach, a new boundary fitted mesh (ALE-FEM1) was
constructed with 14708 elements and $h = 2.5 \text{ cm}$. A step size 
of $\Delta t = 0.01~s$ was used for the temporal discretization.

\begin{figure}[!]
	\centering
	\includegraphics[trim={0cm 5cm 0cm 0cm},clip,width=\textwidth]{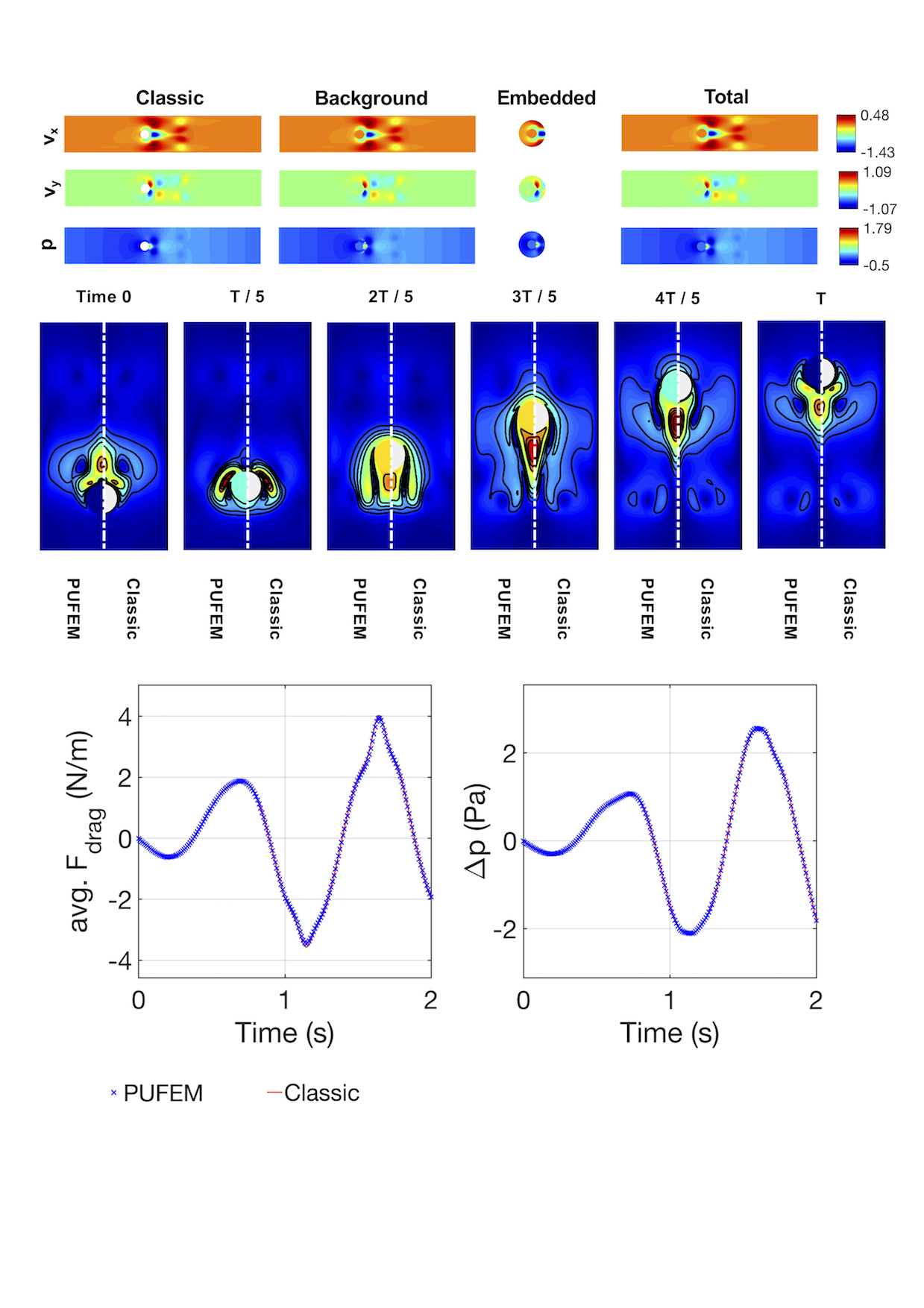}
	\caption{Compiled results for the ALE problem. (Top) A breakdown of the PUFEM solution into background(global), embedded and total. Classical result is introduced for comparison. (Middle) The $x$ component of the velocity as it evolves between the two points of maximum displacement from the centre. Contour lines were added to aid comparison. (Bottom) Estimated drag force and pressure difference across the cylinder for the PUFEM and classical (BF1) methods.}
	\label{fig:ALE_benchmark}
\end{figure}

\begin{figure}[!]
	\centering
	\includegraphics[trim={0cm 10cm 0cm 10cm},clip,width=\textwidth]{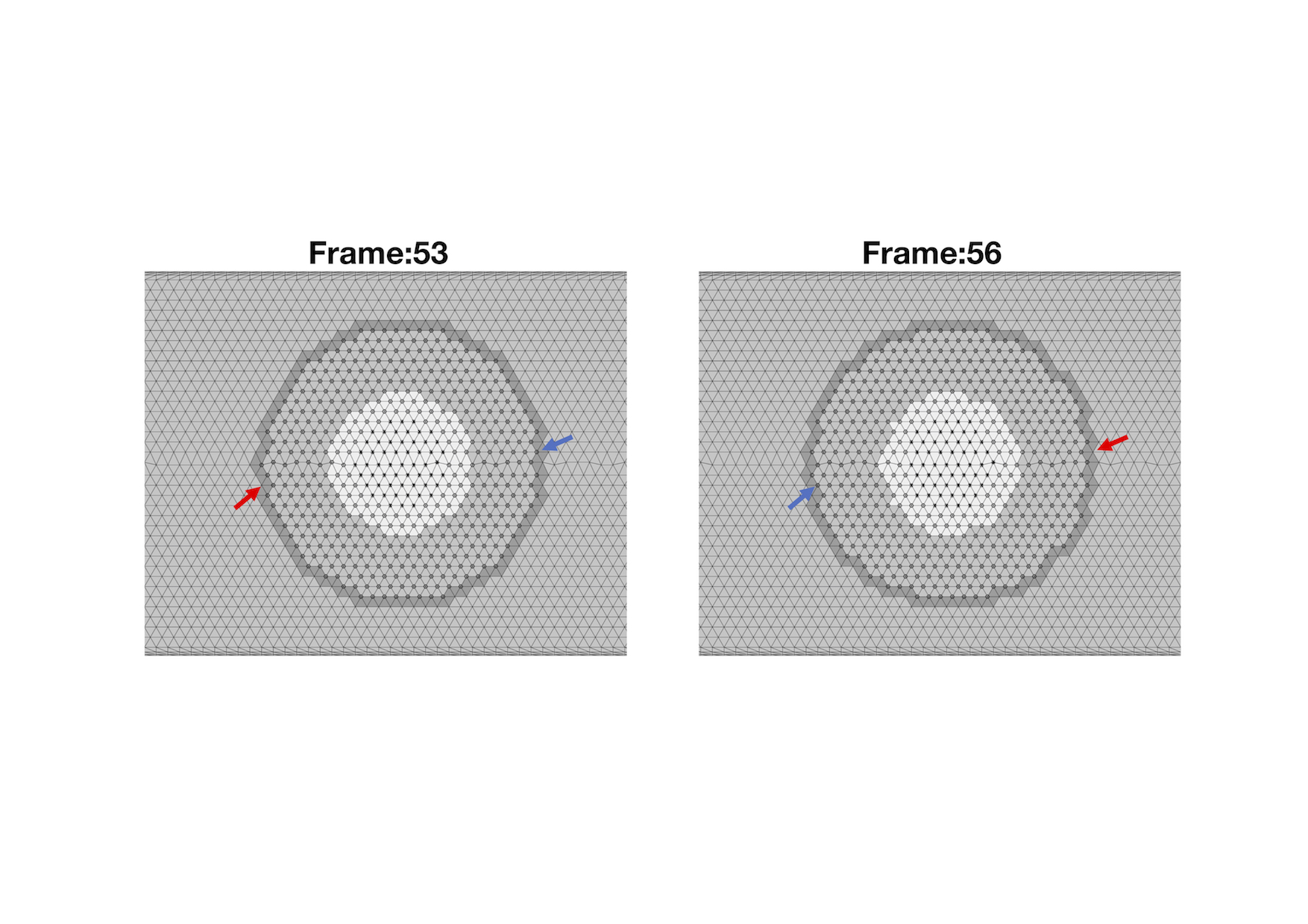}
	\caption{An example of transient fixed nodes on the background mesh. The red arrows point 
	to free nodes, while the blue ones point to their constrained counterpart at different time. The element
	colour coding indicates the following: element cut by $\Gamma^h_\mathit{ff}$ (deep grey), element completely
	weighed out by $\psi^h$ (light grey).}
	\label{fig:ALE_overlap}
\end{figure}

For a qualitative comparison between methods, we maintained the same
format for displaying the flow results as in the case of the
Sch\"{a}fer-Turek benchmark, see Fig.~\ref{fig:ALE_benchmark}.
The top part of the figure is a snapshot
at $t = 2 \hspace{1mm} s$ showing the velocity components and pressure
field as computed on the boundary fitted, the background and embedded
grids. The corresponding PUFEM total field is also included.
 The middle part displays snapshots of the
velocity magnitude at different stages of a transition from right to left. 
In general, the two methods seem to be in agreement, with only minor
differences. 
In contrast to previous examples, the ALE problem is the first case
in this work which includes the concept of transient fixed nodes, see 
Fig.~\ref{fig:ALE_overlap} .Thus,
the good quality of our solution now also suggests the effectiveness
of using the interpolation strategy on providing nodes with a meaningful 
solution while they are temporarily fixed.

We also computed the mean surface drag force ($F_d$) and pressure across the left and
right ends of the cylinder ($\Delta p$). Fig.~\ref{fig:ALE_benchmark} (bottom) shows the
values of these coefficients at all time points of the simulation. The maximum
overall errors that we obtain are $4.0\%$ for the mean $F_d$ and $0.8\%$ for
$\Delta p$. Potential sources which may influence the error (excluding PUFEM)
include mesh deterioration in the case of the classic approach and also the 
fact that we did not aim to minimize the errors through spatial and temporal 
refinement. Despite this, these results suggest that 
PUFEM can be used in an ALE context to obtain reasonably accurate estimations of surface stresses 
and pressure experienced by the solid. 
 This quality is particularly relevant in the case FSI, where the balance of 
tractions is part of the coupling conditions.

\subsection{Valve simulation}
\label{sec:valve-simulation}

\begin{figure}[!]
	\centering
	\includegraphics[trim={0 10cm 0cm 10cm},clip,width=\textwidth]{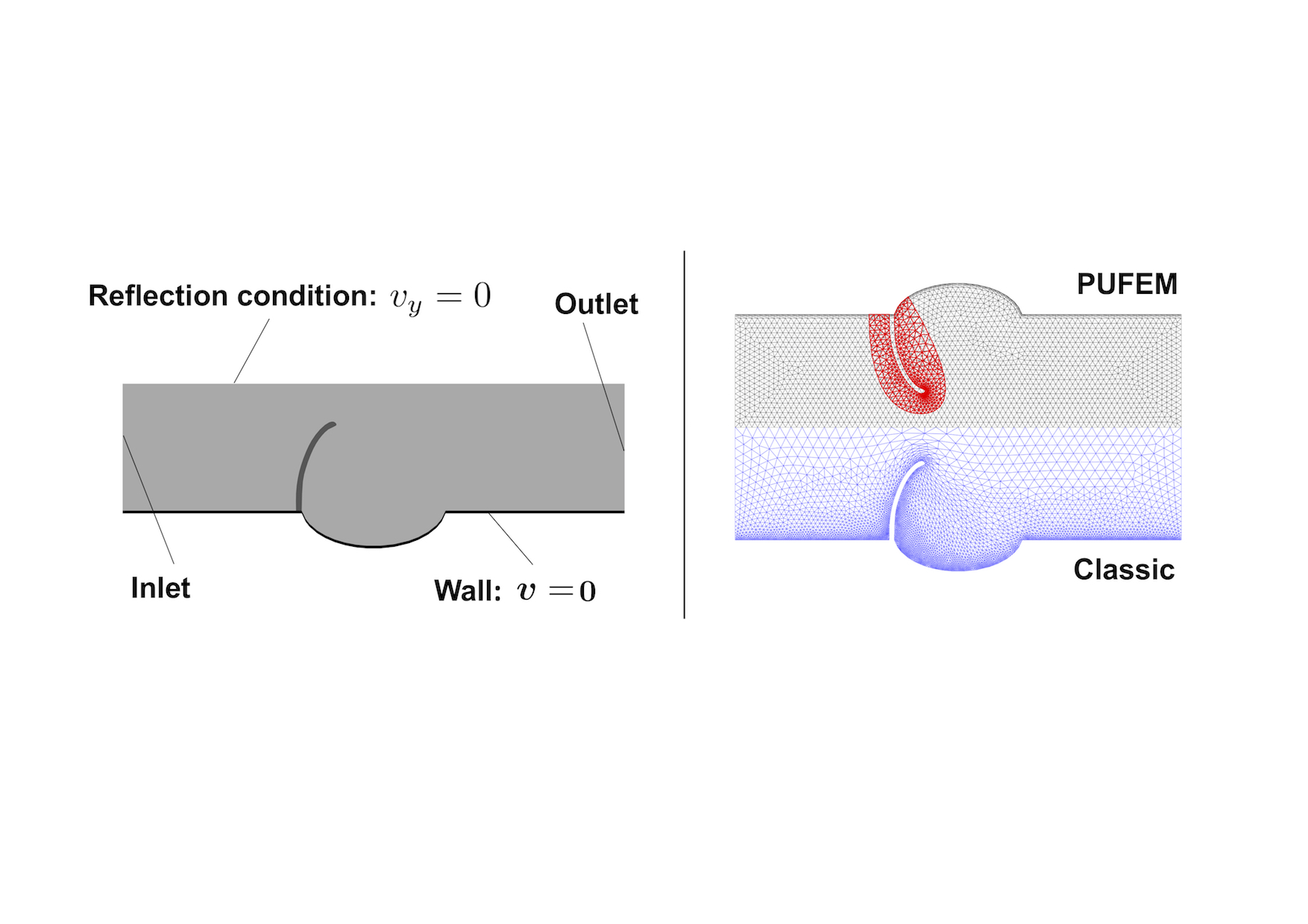}
	\caption{(Left) Illustration of the fluid and valve domains used in the FSI test. A summary 
		of the domain dimensions is found in Table~\ref{tab:FSI_char_params}.
		(Right) The meshes used in the classic and PUFEM approaches. For visualization
                purposes the solid mesh was excluded.}
     \label{fig:valve_domain}
\end{figure}

To examine the suitability of PUFEM for FSI applications, we present
an example of an idealised 2D aortic valve, see Fig.~\ref{fig:valve_domain}.
The FSI domain consists of two components: one half of the lumen and one
leaflet. In order to reduce computation cost, we assume to have rigid
aortic walls and symmetry along the long axis. We also do not include
coronary flow. The fluid is modelled as an incompressible
Navier-Stokes fluid, while the solid is assumed to behave as an
incompressible neo-Hookean material~\cite{hadjicharalambous2014displacement}.
The main domain characteristics
and constitutive law parameters can be found in
Table~\ref{tab:FSI_char_params}.
At the inflow we impose a single pulse with parabolic a
profile:
\begin{equation}
	\bv (y,t) = \left[
		\begin{array}{c}
		V_{max} \frac{y(2W-y)}{W^2} \sin^2(\pi t) \\
		0
		\end{array}\right]
\end{equation}
where $V_{max} = 0.5 \text{ } m/s$ is the maximum velocity, $W$ is the
width of the inflow and $y$ is the spatial coordinate. 
The total in-simulation duration is
$T = 1 \text{ } s$, or one pulse, which we discretise using
constant time steps of $\Delta t = 0.01\text{ } s$. Thus we can observe,
the flow fields both during valve deflection as well as relaxation.
On the surface of the wall and leaflet we impose a no-slip condition, while on
the long axis we apply a symmetry condition. For the outflow, we impose
a zero traction Neumann condition.

In order to reduce the element deterioration and the loss of accuracy which
may derive from it,
the mesh used in the classic approach is built as follows. First, a mesh is
constructed on the reference domain and we run the simulation up to the
$t_{mid} = 0.15 \text{ } s$. The time point was chosen heuristically in order
to capture the configuration of the domain during mid valve deflection. This 
data is then used to construct a second mesh. Using the connectivity matrix of 
this grid, a final mesh is defined by interpolating the $t=0.15 \text{ } s$ node
coordinates onto the space at $t = 0 \text{ } s$. 
 Thus, the classic grid is comprised of 8248 elements, with
mean and maximum $h$ values of $0.4$ mm and $2.8 mm$,
respectively.

For PUFEM, the two grids are generated in the initial domain configuration.
We used a background mesh with 4380 element and an average $h$ of
0.7 mm. The embedded grid is formed of 844 elements with mean $h$ 
of 0.5 mm.

\begin{figure}[!]
	\centering
	\includegraphics[trim={0cm 0cm 0cm 0cm},clip,width=\textwidth]{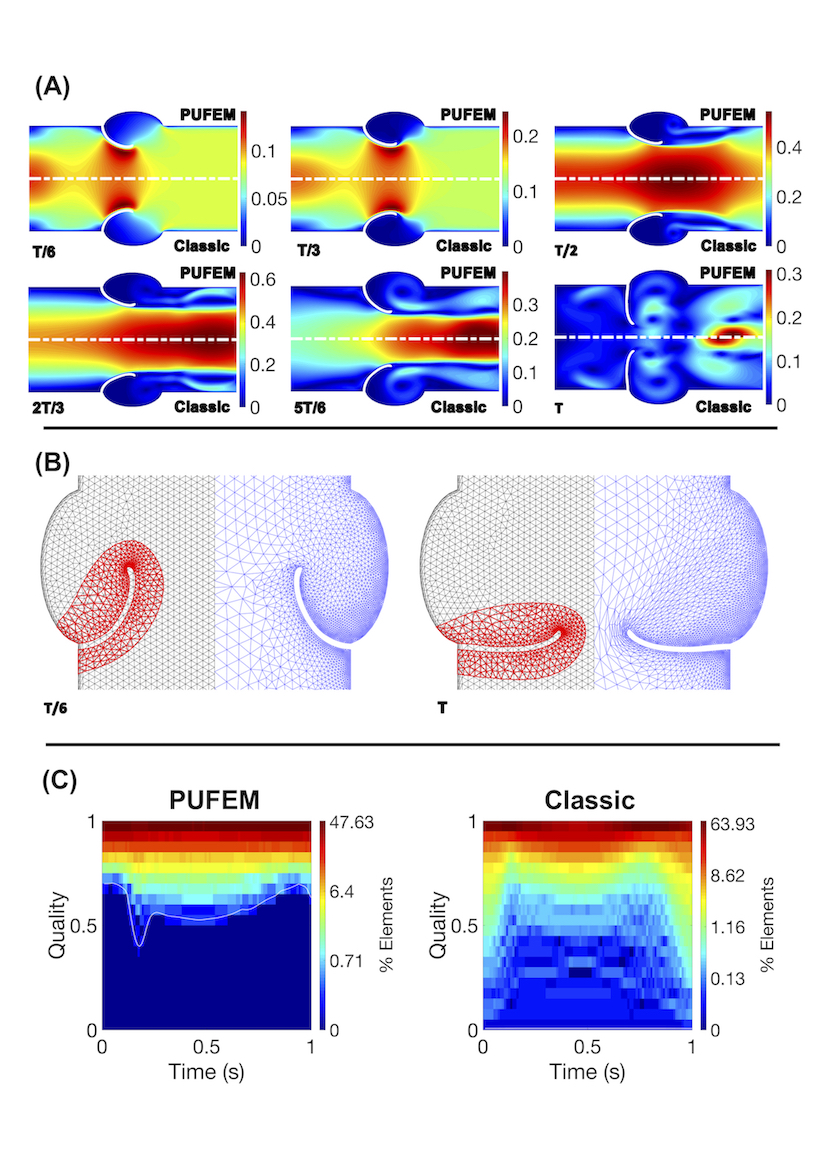}
	\caption{(A) Six snapshots of the velocity magnitude as computed by the PUFEM and classical approaches. (B) The PUFEM and ALE mesh setups at the medium and minimum mesh deflection. At minimum deflection, the ALE mesh reaches a point of almost maximum deterioration. (C) The distribution in time of elements as a function of quality for both approaches.}
	\label{fig:FSI_results1}
\end{figure}

The main flow results are summarized in Fig.~\ref{fig:FSI_results1}: 
the velocity magnitude fields
at different stages of the pulse (A); the grid deformation at half-way
deflection and end time point (B); and the element quality
distribution as it evolves in time (C). 
From a qualitative
perspective, the flow physics appear to be quite similar at
all time stages. While significant differences  appear
in the second part of the simulation, we believe that these are
generally due to a combination of  the 
grid deterioration in the case of the classic 
approach and decelerating flows, which are more sensitive 
to its effects. In fact, without the use midway meshing, to try 
to attenuate these effects, we observed much more striking
difference between methods (results not shown here). The
deterioration of grid quality for the classic approach can be
clearly seen in (B). Thus, while at $T/6$ (midway deflection), 
the boundary fitted mesh typically displays good quality elements,
at $T$ (end of relaxation) many of the elements close to the tip
of the leaflet appear to be squeezed against the axis of symmetry.
On the other hand PUFEM does not show such clear signs,
suggesting that mesh can undergo even more significant 
deformation. This is also reflected in (C), where for PUFEM's
embedded grid the minimum element quality only temporarily
drops bellow 0.5. In the case of the classic approach, the element 
population is selected from the interval $[1.5, 3.7]$ cm on the $x$
axis, as shown in (B). Here we see that while a majority number of elements
retains an excellent quality, for many the value is very low, close to 
collapsing. The reason for having a much wider population spread
at the start and end of the simulation is a result of using the midway
meshing strategy. In return, the population appears to be much 
more skewed towards one in the middle half of the simulation,
when the we experience the highest inflow velocity.

\begin{figure}[!]
	\centering
	\includegraphics[trim={0 0cm 0 0cm},clip,width=\textwidth]{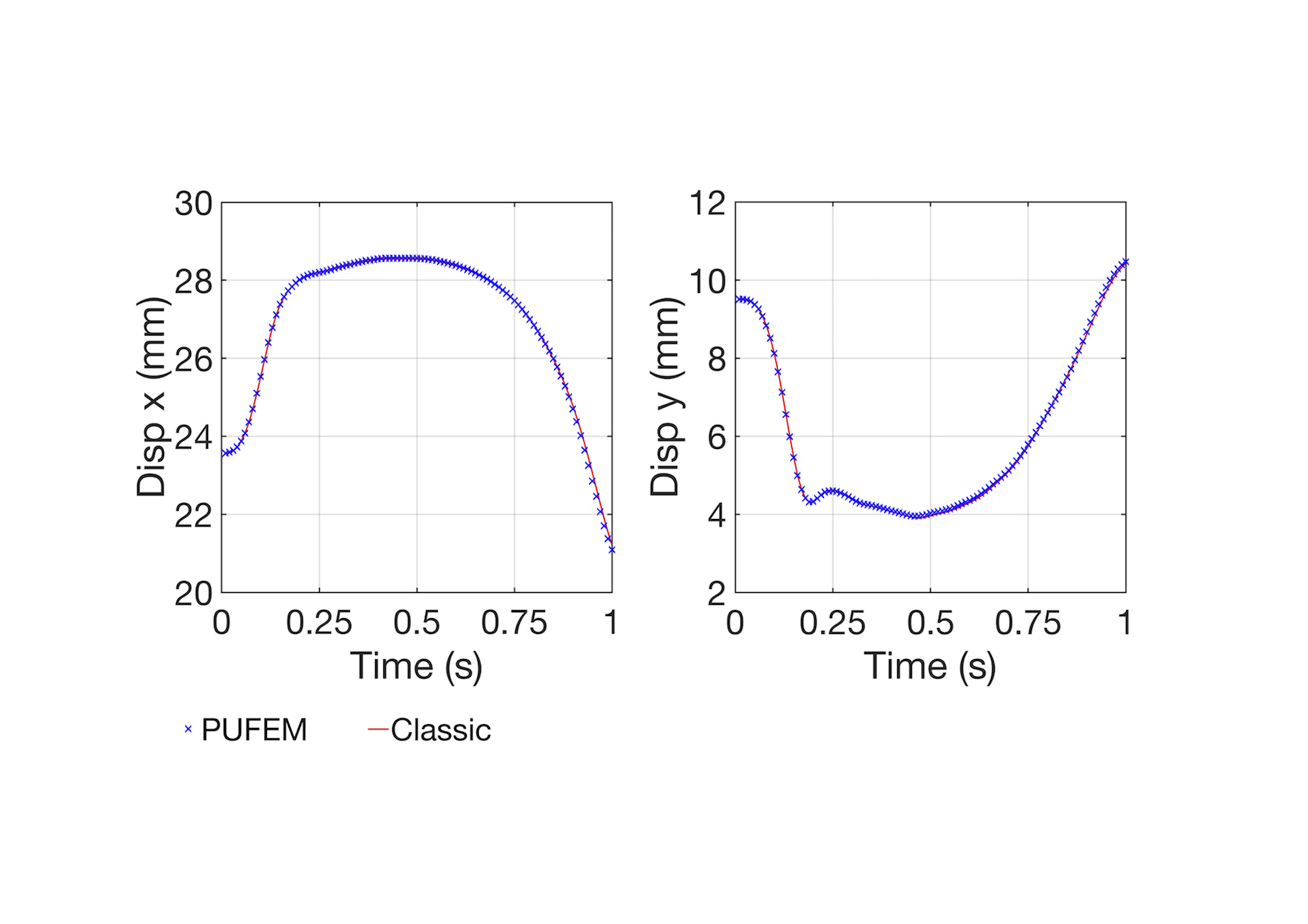}
	\caption{The $x$ and $y$ displacements of the valve tip as it evolved in the classic and PUFEM approaches. 
    The reference nodes where chosen such that they have identical coordinates at $t = 0~s$.}
	\label{fig:FSI_results2}
\end{figure}

From the solid mechanics perspective, a good agreement can also be
observed for the valve tip deflection, see Fig.~\ref{fig:FSI_results2}, with
maximum absolute errors of $0.21$ and $0.1 \text{ } mm$ in $x$ and $y$
directions, respectively.

Combined, these preliminary results on FSI suggest not only
show a good level of accuracy for PUFEM, comparable to that 
of the classic approach, but also to a greater degree of flexibility,
where we can better retain element quality while undergoing 
significant mesh deformation.

%%% Local Variables:
%%% mode: latex
%%% TeX-master: "Max_Article"
%%% End:

\section{Conclusion}
\label{sec:conclusion}
In this work we have presented a novel approach to solving fluid flow
and FSI problems based on the partition of unity (PUFEM). Thus, the total
flow solution is defined as weighted sum of background and embedded 
fields. The weighting field, which is defined on the embedded mesh,
is used to facilitate a smooth transition between areas where the solution
is dominated by one mesh or the other.

The accuracy and stability of the method has been assessed on a range of 2D tests with
an increasing level complexity. For comparison, all tests were replicated using a 
classic boundary fitted FEM approach which served as our golden standard. Using
steady-state results for flow around the cylinder, we showed that PUFEM displayed
nearly optimal convergence rates for both Stokes and Navier-Stokes, and that the
the accuracy of the two approaches is nearly indistinguishable for similar
spatial discretization. This  suggests that PUFEM can leverage the approximation power of 
both meshes with little deterioration as a result of the coupling process.

In the case of the Sch\"afer-Turek, the flow results produced by PUFEM where shown 
to be in good agreement with the classic approach. Additionally, 
the estimation of surface stresses, pressure and wake frequency
 appeared to be in accord with literature values.

To assess the accuracy and stability of PUFEM in the ALE setting, we designed a problem
where flow is generated in an open-ended channel by an oscillating, rigid cylinder. The method
appeared robust to the changing of the overlapping area and the set of constrained nodes.
Additionally, a good agreement was maintained between approaches in the estimation of 
surface drag forces and pressure drop.

An idealised aortic valve FSI problem was chosen for testing of realistic applications. Given
the set up the problem, the resulting flow is complex with the occurrence of recirculation 
and small vortexes during relaxation. In addition to displaying a high level of similarity 
between the physical behaviour obtained using the two approaches, we where able
to gauge the evolution of mesh quality over the course of the simulation. Thus, 
the use of overlapping meshes appears to produce the desired effect of reducing 
the well known mesh degradation associated with large deformations.

Combined, these results indicate the potential applicability of PUFEM to a wide 
range of FSI applications, allowing for boundary fitted fluid-solid interface and
 targeted solution space enrichment, in the absence of the need for re-meshing
 or the need for user-defined stabilization parameters.

Future work will focus on providing a theoretical analysis on the stability and
convergence of the PUFEM solver. Other extensions of this work will include
porting it into a parallel 3D implementation~\cite{lee2016multiphysics} and 
the study of different approaches to include contact 
mechanics.

%\section*{Conclusions}
%In this work we have presented our new approach for solving fluid
% flow using PUFEM-based overlapping domain technique. While the
% method can potentially be applied to a wide range of problems, here
% we covered a couple of important examples, namely the Stokes problem
% and the ALE form of the Navier-Stokes equation. Additionally, we
% also proposed a set-up for integrating the method into a monolithic
% FSI solver.

% Based on our tests, we have shown that the method is accurate and
% stable for steady-state Stokes and Navier-Stokes equations. The
% method also displays good agreement with the 2D Turek benchmark as
% well as out own ALE benchmark. Finally, for more practical
% applications, we show our preliminary results in the case of a 2D
% FSI problem which simulation the flow of blood past an aortic valve.

% In future work we aim to provide a mathematical analysis of the our
% PUFEM-based approach. Furthermore, we will try to address one of the
% more challenging problems of overlapping domain techniques, which is
% how to to incorporate contact problems into the FSI set-up.

\section*{Acknowledgments}
D.N. acknowledges funding form the Engineering and Physical
Sciences (EP/N011554/1 and EP/R003866/1).
A.M. gratefully acknowledges financial support from the Swedish
Research Council under Starting Grant 2017-05038 and from the
Wenner-Gren foundation under travel grant SSh2017-0013.
JH acknowledges the financial support of the Swedish Research Council 
under Grant 2018-04854.
This work is funded by the King's College London and Imperial
College London EPSRC Centre for Doctoral Training in Medical 
Imaging (EP/L015226/1). This work is supported by the Wellcome
EPSRC Centre for Medical Engineering at King's College London
(WT 203148/Z/16/Z) and by the National Institute for Health Research
(NIHR) Biomedical Research Centre award to Guy and St Thomas'
NHS Foundation Trust in partnership with King's College London.
The views expressed are those of the authors and not necessarily
those of the NHS, the NIHR or the Department of Health.

\section*{Additional Information}
Declarations of interest: none.

\newpage

% A reference header is automatically set/printed by the elsevier style
%\section*{References}

\section*{References}
\bibliography{mybibfile}

\newpage

\newpage

\newpage

\newpage

\newpage

\begin{table}
	\tiny
	\begin{tabularx}{\textwidth}{X | X X X X X }
		\hline \hline
		& Level 1 & Level 2 & Level  3 & Level 4 & Level 5 
		\\ \hline \hline
		Classic  \\ \hline
		h & 0.1 & 0.05 & 0.025 & 0.0125 & $0.1/2^4$  \\
		element no. & 225 & 880 & 3488 & 13774 & 54746 \\
		nodes (P1) & 136 & 488 & 1840 & 7079 & 27757 
		\\ \hline \hline
		M1 &  Level 1 & Level 2 & Level  3 & Level 4 & Level 5 \\
		\hline
	    $h$ & 0.1 & 0.05 & 0.025 & 0.0125 & $0.1/2^4$  \\
	    B. element no. & 232 & 926 & 3704 & 14816 & 59264 \\ 
	    E. element no. & 76 & 296 & 1196 & 4454 & 17310 \\
	    B. nodes (P1) & 137 & 504 & 1933 & 7569 & 29953 \\ 
	    E. nodes (P1) & 48 & 166 & 630 & 2295 & 8789
		\\ \hline \hline
		M2 \\
		\hline
		Background $h$ & 0.1 & 0.05 & 0.025 & 0.0125 & $0.1/2^4$  \\
		Embedded $h$ & 0.05 & 0.025 & 0.0125 & $0.1/2^4$ & $0.1/2^5$ \\
		E. element no. & 296 & 1196 & 4454 & 17310 & 69110 \\
		E. nodes (P1) & 166 & 630 & 2295 & 8789 &  34822
		\\ \hline
	\end{tabularx}	
	\caption{Statistics corresponding to the mesh sets used in the steady Stokes and Navier-Stokes convergence tests.}
	\label{tab:mesh_stat}
\end{table}

\newpage

\newpage

\newpage

\newpage

\newpage

\newpage

\begin{table}[h]
	\tiny
	\begin{tabularx}{\textwidth}{ X |X| X X |X X}
		\hline \hline
		& FEM & \multicolumn{4}{c}{PUFEM} \\
		\hline \hline
		Label& FEM1 &  \multicolumn{2}{l|}{PU1}      &  \multicolumn{2}{l}{PU2}  \\
		&           &  \multicolumn{1}{l}{Background}        &  \multicolumn{1}{l|}{Embedded}   & \multicolumn{1}{l}{Background}          &  \multicolumn{1}{l}{Embedded}\\
		\hline 
		Elements & 14684 & 14816 & 1106 & 56362 & 4390  \\
		DOF       & 67208 & 67745 & 5189 & 255742 & 20171 \\
		${h} $ (cm)  & 2.5 & 2.5 & 2.5 & 1.25  & 1.25 \\
		\hline
	\end{tabularx}
	\caption{Number of elements, DOF and average element size for the meshes used in the Turek benchmark.}
	\label{tab:mesh_turek}
\end{table}

\newpage

\begin{table}[!]
	\tiny
	\begin{tabularx}{\textwidth}{ X| X X X X X}
		\hline \hline
		& Time steps & $c_d$ & $c_l$ & $St$ & $\Delta p$ \\
		\hline 
		Lit.~\cite{schafer1996benchmark} & - & $3.22 - 3.24$ & $0.99 - 1.01$  & $0.295 - 0.305$ & $2.46 - 2.50$ \\
		FEM1  & 1000 &  $3.054$ & $ 0.906 $ & $0.2941$  & $2.4950$   \\
		PU1    & 1000 &  $3.053$ & $ 0.904 $ & $0.3030$ & $2.4911$  \\
		PU1    & 2000 & $3.060$ & $ 0.926 $ & $0.3030$ & $2.4981$ \\
		PU1    & 5000 & $3.061$ & $ 0.929$ & $0.3030$ &  $2.4988$ \\
		PU2   & 1000 & $3.175$  & $0.950$ & $0.2941$ & $2.4801$ \\
		PU2   & 2000 & $3.179$  & $0.970$ & $0.3030$ & $2.4870$ \\
		PU2   & 5000 & $3.182$ & $0.976$  & $0.3030$ & $2.4820$ \\
		\hline
	\end{tabularx}
	\caption{Coefficient estimates from PUFEM and classic FEM for the Turek benchmark.}
	\label{tab:coeff_turek}
\end{table}

\newpage

\newpage

\newpage

\newpage

\begin{table}[!]
	\centering
	\tiny
	\begin{tabularx}{\textwidth }{X X X X X X X}
		\hline \hline
		\multicolumn{2}{l}{\textbf{Domain dimensions}} \\
		\hline
		&  Inflow width & Max width & Aortic length & Sinus length & Valve height  \\
		Distance (mm) & 14 & 17.9 & 55.3 & 15.8 & 9.8  
		\\ \hline \hline
		\multicolumn{3}{l}{\textbf{Fluid parameters}}  & 
		\multicolumn{3}{l}{\textbf{Solid parameters}}  \\
		\hline
		$\mu_f \text{ } (Pa\cdot s)$ & $3\times10^{-3} $& & $\mu_s \text{ } (Pa\cdot s)$ & $20\times 10^{3}$ \\
		$\rho \text{ } (kg/m^3)$ & 1030 \\
		\hline
	\end{tabularx}
	\caption{(Top) The main domain dimensions in the aortic valve problem. 
		(Bottom) The constitutive law parameters for the fluid and solid problems.}
	\label{tab:FSI_char_params}
\end{table}

\newpage

\newpage

\newpage

\end{document}